%% file: Main.tex
\documentclass[acmtog,nonacm=true]{acmart}

\usepackage{amssymb}
\usepackage{subcaption}
\usepackage{siunitx}

\newcommand{\R}{\mathbb{R}}
\newcommand{\had}{\odot}                    
\newcommand{\E}{\mathrm{E}}                 
\newcommand{\D}{\mathrm{D}}                 
\newcommand{\secref}[1]{Sec.~\ref{#1}}
\title[Learned Spectral Codes for RGB Pipelines]%
      {Spectral Rendering Without a Spectral Renderer:\texorpdfstring{\\}{ }
       Learned Spectral Codes for RGB Pipelines}

\author{Jiaqi Yu}
\email{jiaqiyu0307@outlook.com}
\orcid{0009-0002-6168-2204}
\affiliation{%
  \institution{University of York}
  \city{York}
  \country{United Kingdom}}

\author{Dar'ya Guarnera}
\email{darya.guarnera@york.ac.uk}
\orcid{0000-0001-8751-5215}
\affiliation{%
  \institution{University of York}
  \city{York}
  \country{United Kingdom}}

\author{Giuseppe Claudio Guarnera}
\email{claudio.guarnera@york.ac.uk}
\orcid{0000-0002-7703-5194}
\affiliation{%
  \institution{University of York}
  \city{York}
  \country{United Kingdom}}
\affiliation{%
  \institution{Lumirithmic Ltd}
  \city{London}
  \country{United Kingdom}}

\begin{document}

\begin{teaserfigure}
  \centering
  \includegraphics[width=\linewidth]{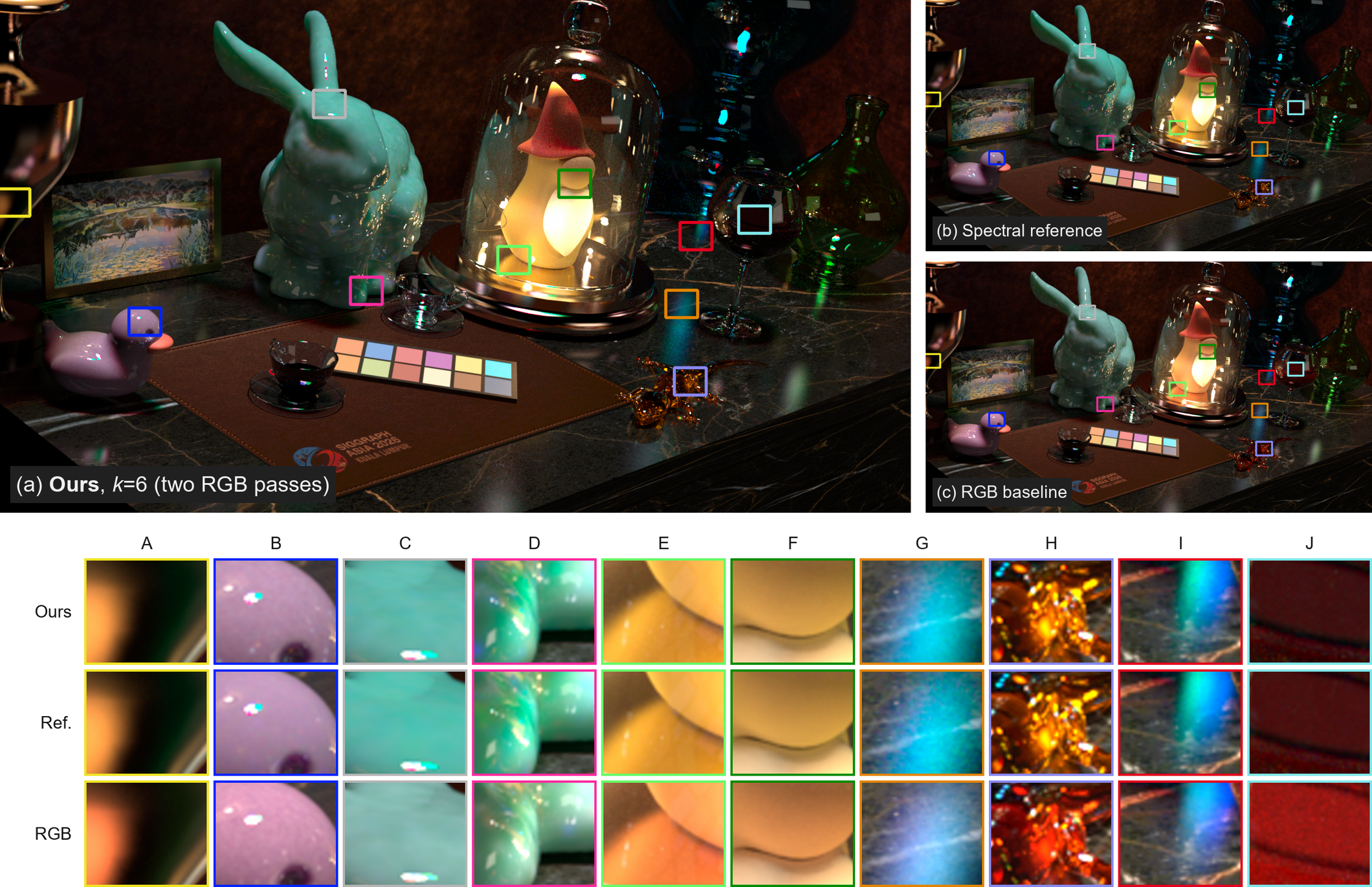}
 \captionsetup{skip=3pt}
 \caption{\textbf{A spectral image rendered through an unmodified RGB path tracer.} Our codec (a) approximates the spectral ground truth (b) in two standard RGB passes; the conventional RGB render (c) shows visible chromatic shifts. The insets (A)--(J) magnify the regions marked in (a)--(c), rendered at higher sample counts; the three rows of each inset column share one exposure, normalised per patch for legibility. The scene exercises every supported material class (spectral textures, measured Schott glass and clear glass, copper and gold conductors, dipole and homogeneous subsurface scattering, and participating-medium fog) under mixed warm/cool broadband illumination with red, green, and blue accents.}
\label{fig:teaser}
\end{teaserfigure}

\begin{abstract}
Spectral rendering reproduces the wavelength-dependent appearance
that RGB rendering cannot: metamerism, colour shifts under
spectrally rich illumination, coloured glass, metallic reflection
and subsurface scattering. It requires a spectrally aware renderer,
but the rendering ecosystem is predominantly RGB: game engines,
cloud render farms, GPU rasterisers and most offline renderers
consume RGB alone, with no access to the internal wavelength
propagation that spectral rendering relies on.

We introduce \emph{learned spectral codes}, a compact linear
representation that enables general spectral rendering through
\emph{unmodified} RGB pipelines.  A trained encoder compresses any
visible-range spectral quantity into a small latent vector; the
codes are packed as RGB triplets and processed by an off-the-shelf
RGB renderer.  The codec preserves the arithmetic of light
transport, exactly for scaling and addition and approximately for
element-wise products, so two standard RGB passes suffice to
produce results closely matching a full spectral rendering.  The
same codec handles reflectance, illumination, coloured glass,
conductors, and participating media: one pipeline across material
classes, with no per-material modification to the renderer.

Where spectral asset data is unavailable, lightweight neural
upsamplers bridge legacy RGB inputs to the latent space.
Across diverse materials and illumination, the method yields
substantial quality improvements on the cases where RGB rendering
fails most visibly, at rendering cost comparable to efficient
spectral-sampling techniques and with no changes to the renderer.
\end{abstract}



\setcopyright{cc}
\setcctype{by}
\acmJournal{TOG}
\acmYear{2026} \acmVolume{45} \acmNumber{6} \acmArticle{257}
\acmMonth{12} \acmDOI{10.1145/3842506}

\maketitle

\input{sections/introduction.inc}

\input{sections/related_work.inc}

\input{sections/codec.inc}

\input{sections/material_encoding.inc}

\input{sections/upsampling.inc}
\input{sections/experiments.inc}

\input{sections/conclusion.inc}

\begin{acks}
This work was partly supported by the Arts and Humanities Research
Council (AHRC) grant AH/Y001079/1 (CoSTAR Live Lab).
\end{acks}

\bibliographystyle{ACM-Reference-Format}
\bibliography{bibliography}


\end{document}

%% file: sections/introduction.inc
\section{Introduction}

Spectral rendering propagates full spectral distributions through the
rendering pipeline, reproducing wavelength-dependent phenomena that
RGB rendering cannot: metamerism, prismatic dispersion, and thin-film
interference~\cite{Watanabe2013}. Modern
spectral renderers achieve this efficiently: hero wavelength
sampling~\cite{Wilkie2014} and GPU-accelerated spectral path
tracers~\cite{murray2021efficient} bring spectral rendering to near-RGB
cost in open-source and custom production pipelines.

Much of the wider rendering ecosystem, however, remains RGB-based.
Spectral pipelines such as Manuka~\cite{Fascione2018} exist in
production, but commercial game engines, open-source suites such as Blender,
GPU rasterisers, cloud render farms, and most proprietary offline
renderers expose RGB inputs and RGB framebuffers only; modifying
them to propagate stochastic wavelengths is impractical.

Recent work delivers individual spectral effects without touching
the renderer core: Tominaga et al.~\shortcite{Tominaga2026} render
fluorescence through Mitsuba\,0.6's spectral build by encoding the
re-radiation into per-channel inputs, and Fichet et
al.~\shortcite{fichet2024fluorescence} obtain a similar effect with
a reduced re-radiation matrix implemented at the shader level.
Fluorescence itself lies outside our supported regime, so these
methods are complementary to ours.

Rendering general spectral content through multiple passes of an
RGB pipeline originates with Darling et
al.~\shortcite{Darling2011}, who relight scenes with a fixed
six-channel parametric basis. We generalise this assets-only route,
which requires no custom code anywhere: a single representation for
per-channel-preserving spectral transport across reflectance,
illumination, glass, conductors, and participating media, rendered
through entirely unmodified RGB renderers. This representation, a
\emph{learned spectral code}, is a compact latent vector $z\in\R^k$
that supports the per-channel arithmetic an RGB renderer already
performs: scaling, addition, and element-wise multiplication. Spectral quantities encode into $z$;
the codes pack into $k/3$ RGB triplets; the renderer processes them in $k/3$ unmodified passes; a final decode recovers the spectral radiance (Fig.~\ref{fig:overview} illustrates the full pipeline). Whether the renderer is Mitsuba, Blender Cycles, Maya Arnold, KeyShot, or a real-time rasteriser (Unreal Engine 5) is immaterial: each processes RGB triplets per channel, which is what the codec
requires.

Exact preservation of all three operations is impossible whenever the
code dimension $k$ is smaller than the number $n$ of spectral bins
(\secref{sec:impossible}), so the encoder is approximate by necessity.
Our linear, non-negative architecture preserves scaling and addition
\emph{exactly} and is trained to approximate element-wise multiplication
on the manifold of natural visible spectra. Since each RGB pass
carries three channels, practical operating points are multiples of
three; \secref{sec:capacity_choice} establishes $k{=}6$ as the
smallest configuration that is meaningfully more accurate than a
standard RGB pipeline, with $k{=}3$ blocked by a Luther--Hadamard
trade-off (alignment with the colour-matching
subspace~\cite{Ives1915,Luther1927} against Hadamard preservation)
and $k{=}9$ unwarranted in quality terms.

The codec is content-agnostic: the same encoder--decoder embeds
reflectance, illumination, transmittance, conductor IOR, and
absorption and scattering coefficients into the same latent space
(\secref{sec:materials}); two per-pixel upsamplers project legacy
RGB assets onto the codec's manifold (\secref{sec:upsampling}).
Our scope covers wavelength-preserving interactions;
wavelength-shifting effects (fluorescence, dispersion) lie
outside the supported regime (\secref{sec:conclusion}).

\noindent\textbf{Contributions.}
\begin{enumerate}
  \item A \textbf{learned non-negative spectral basis and its dual},
        jointly optimised so that iterated per-channel products of
        the resulting $k$-dimensional codes remain accurate. This
        codec compresses any visible-range spectral quantity and
        enables general spectral rendering through unmodified RGB
        renderers in $k/3$ standard passes (\secref{sec:materials}).
  \item \textbf{Latent upsamplers} for legacy RGB assets that
        target the codec's $k$-dimensional manifold rather than
        full $n$-channel spectra (\secref{sec:upsampling}).
  \item A \textbf{capacity analysis} establishing $k{=}6$ as the
        practical operating point: $k{=}3$ blocked by a
        Luther--Hadamard trade-off, $k{=}9$ unwarranted in quality
        terms, and the learned basis irreplaceable at $k{=}6$
        (\secref{sec:capacity_choice}).
\end{enumerate}

We evaluate the codec on broadband, narrowband, and multi-peaked illumination across surface reflectance, coloured-glass transmission, conductor reflection, and homogeneous and heterogeneous subsurface scattering; the teaser (Fig.~\ref{fig:teaser}) exercises all of these in one scene. We benchmark timing and reconstruction quality against Mitsuba\,3's hero wavelength sampling at matched resolution and spp, and we deploy the codec across four production renderers (\secref{sec:experiments}).

%% file: sections/related_work.inc
\section{Related Work}
\label{sec:background}

\subsection{Compact spectral representations}
\label{sec:rw:representations}

Most compact spectral representations, ours included, are linear
decompositions.
Peercy~\shortcite{Peercy1993} introduced low-dimensional
point-sampled representations for spectral rendering; PCA has 
been applied to reflectance and illumination
datasets~\cite{Otsu2018} and to measured material
capture~\cite{Matusik2003, Guarnera2017}.  PCA admits negative
entries; NMF~\cite{Berry2007} is non-negative but needs more
dimensions.  Both serve as baselines in
\secref{sec:capacity_choice}.

Parametric representations replace the data-fitted basis with a
constrained family. Peters et al.~\shortcite{Peters2019} encode bounded
spectra as Fourier moments with a maximum-entropy reconstruction,
achieving 3--8 coefficients with smooth, energy-conserving output.
Karimipour et al.~\shortcite{Karimipour2023} represent reflectance as
$3\!\times\!3$ matrices fitted to cone responses, reaching
$\Delta E_{94}\!<\!1$ under broadband illumination but degrading to
$\Delta E_{94}$ up to 16 under narrowband sources. Both compress
\emph{individual spectra} accurately but provide no algebra on the
compressed codes: the product of two compressed spectra cannot be
evaluated without decompressing first.

\subsection{Efficient transport in spectral renderers}
\label{sec:rw:transport}

Hero wavelength sampling~\cite{Wilkie2014} carries one stochastically chosen wavelength per ray with several stratified companions, reducing colour noise at near-RGB cost. West et al.~\shortcite{West2020} generalise wavelength importance sampling to a continuous multiple importance sampling framework, and van de Ruit and Eisemann~\shortcite{vandeRuit2021} build coarse screen-space estimates over multiple passes to guide subsequent sampling. These methods improve efficiency \emph{within} a spectral renderer; we target renderers with no spectral path at all.

\subsection{RGB-to-spectrum upsampling}
\label{sec:rw:uplift}

Converting RGB textures to plausible spectra is a long-standing
problem. Smits~\shortcite{Smits1999} introduced a scaled
basis-spectrum conversion with a long production history: the Manuka
renderer relied on a Smits-style optimisation for all its
uplifting~\cite{Fascione2018}, and recent versions replace it with
the moment-based method of Peters et al.~\shortcite{Peters2019} for
reflectances~\cite{Weidlich2021}. Meng et
al.~\shortcite{Meng2015} introduced a physically
meaningful uplift over the XYZ space. Jakob and
Hanika~\shortcite{Jakob2019spectral} parameterise the sRGB gamut with
three coefficients, smooth and energy-conserving; the method is now
standard in PBRT-v4~\cite{Pharr2023} and
Mitsuba\,3~\cite{Mitsuba3}. Jung et al.~\shortcite{Jung2019}
uplift gamuts wider than sRGB by adding fluorescent components.
Belcour et al.~\shortcite{Belcour2023} give metameric one-to-many
uplift for textured assets. T\'odov\'a
et al.~\shortcite{Todova2021, Todova2022} constrain the uplift so that
selected spectra are reproduced exactly; T\'odov\'a and
Wilkie~\shortcite{Todova2025} extend uplifting to HDR environment
maps, and van de Ruit and Eisemann~\shortcite{vandeRuit2024}
constrain uplifted spectra by their rendered RGB appearance,
including indirect illumination. Guarnera et
al.~\shortcite{Guarnera2022}
target illumination spectra specifically, uplifting a light's
tristimulus value by optimising blending weights over a compact
spectral basis. They propose two variants: a measured basis of six
LED primaries, which produces SPDs with the spectral structure of
real multi-primary sources, and a smooth parametric Gaussian basis.
Both reproduce the target chromaticity by construction; as with
reflectance uplifting, the reconstructed SPD is one metamer of many,
and the choice of basis decides which one. Xia et al.~\shortcite{Xia2025}
sample the null space of the RGB constraint via differentiable
rendering. We benchmark our reflectance upsampler against Jakob and
Hanika~\shortcite{Jakob2019spectral} and our illumination upsampler
against Guarnera et al.~\shortcite{Guarnera2022} in
\secref{sec:results_upsampling}. While these methods uplift RGB to
$n$-channel spectra rendered with a spectral renderer, ours map
RGB to $k$-dimensional latent codes rendered with an RGB renderer.

\subsection{Spectral effects in RGB-only renderers}
\label{sec:rw:rgb-engines}

A small but growing line of work delivers spectral phenomena without
a spectral renderer. Ward and
Eydelberg-Vileshin~\shortcite{Ward2002} prefilter materials and lights
before rendering, achieving near-exact direct illumination at the
cost of approximate interreflection. Sun~\shortcite{Sun2006} decomposes
thin-film iridescence into a small number of RGB passes through an
unmodified renderer. Fichet et al.~\shortcite{fichet2024fluorescence}
handle fluorescence by replacing the albedo--light product with a
reduced re-radiation matrix, requiring shader-level changes; their
reduced-matrix formalism itself is general, and we adopt it in
\secref{sec:fichet_view} to analyse our codec's constraints.
Each of these methods is constructed for a
specific phenomenon. We take a different route: a single learned
codec supports general \emph{wavelength-preserving} spectral
transport (reflectance, illumination, coloured glass, conductors,
and homogeneous and heterogeneous participating media) through any
RGB renderer with no engine modification.
Wavelength-shifting effects (fluorescence, phosphorescence,
dispersion) couple distinct wavelengths during transport and remain
outside the per-channel regime our codec preserves.

%% file: sections/codec.inc
\section{Learned Spectral Codes}
\label{sec:method}

\begin{figure*}[t]
  \centering
  \includegraphics[width=0.49\linewidth]{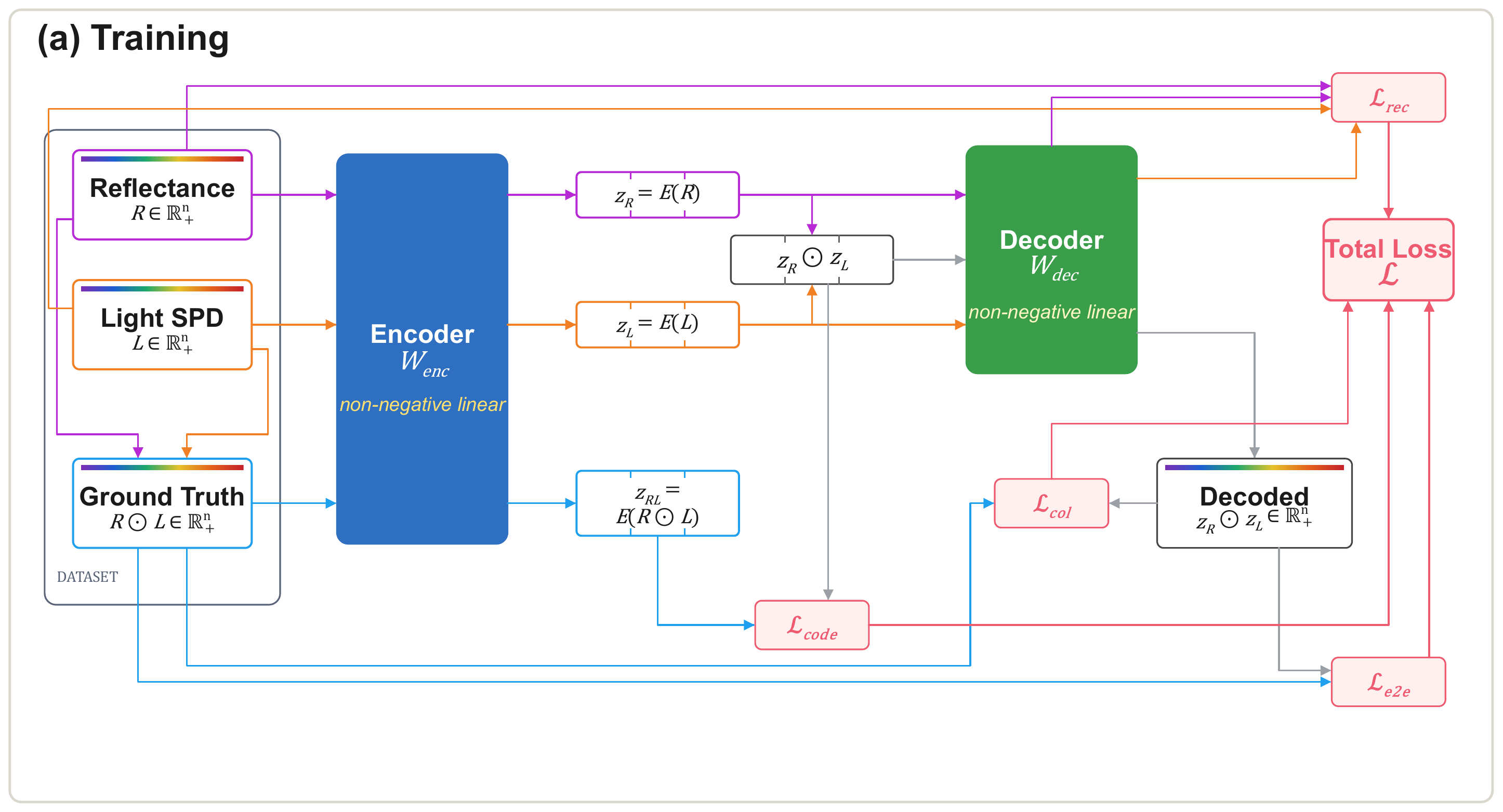}\hfill
  \includegraphics[width=0.49\linewidth]{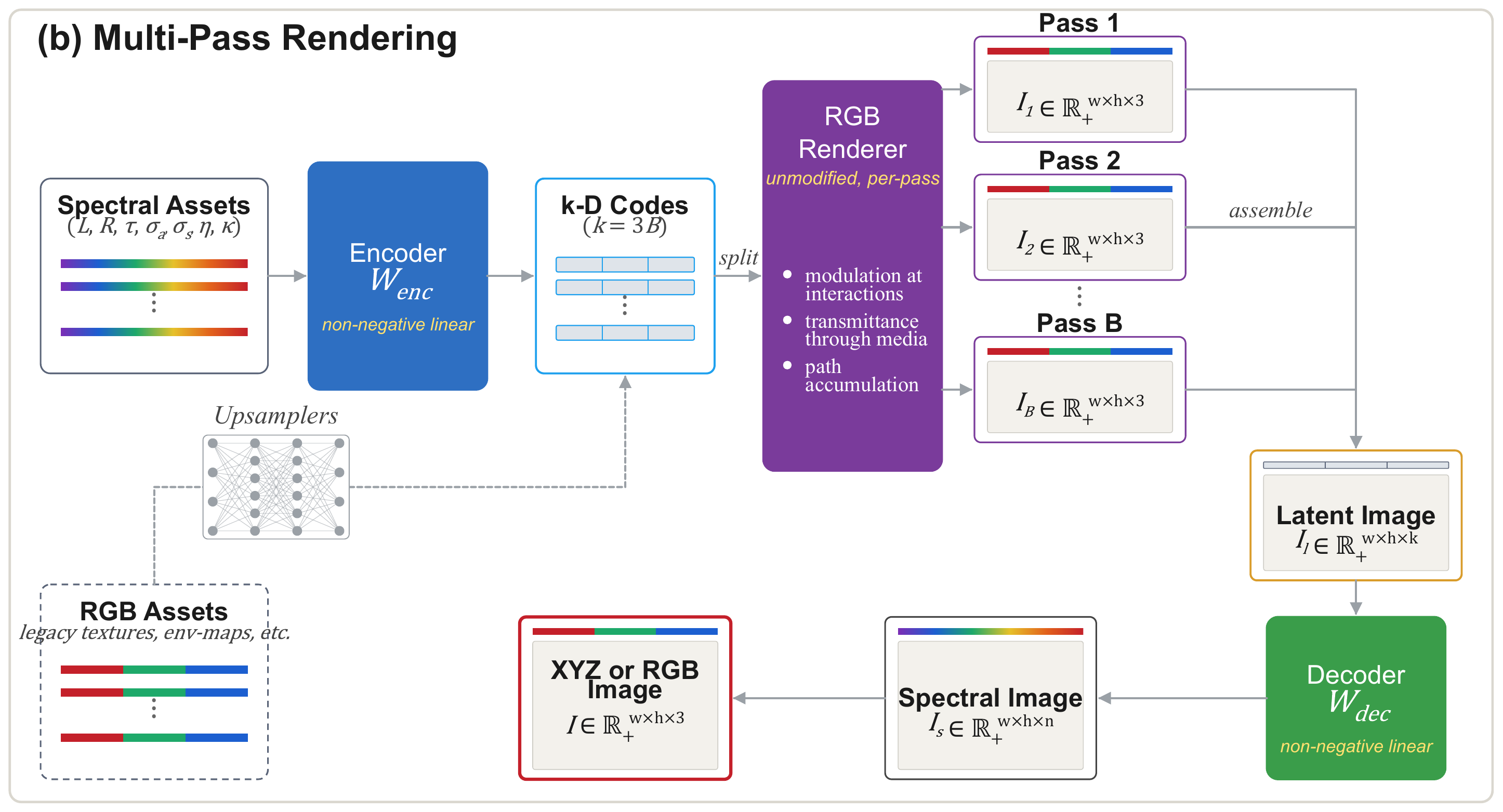}
  \caption{\textbf{Learned spectral codec and multi-pass rendering
pipeline.} \textbf{(a) Training:} a non-negative encoder $\E$
($W_\mathrm{enc}\!\in\!\R_+^{k\times n}$) and decoder $\D$
($W_\mathrm{dec}\!\in\!\R_+^{n\times k}$) are fit jointly on
reflectance and illumination spectra; the same matrices embed
transmittance $\tau$, absorption $\sigma_a$, scattering $\sigma_s$,
and conductor IOR $\eta,\kappa$ at inference. Four loss terms
($\mathcal{L}_\mathrm{e2e}$, $\mathcal{L}_\mathrm{rec}$,
$\mathcal{L}_\mathrm{code}$, $\mathcal{L}_\mathrm{col}$) are combined
(\secref{sec:losses}). \textbf{(b) Rendering:} spectral assets encode
once into $k$-dimensional codes ($k{=}3B$); legacy RGB assets bypass the
encoder via per-pixel upsamplers $f^{(R)},f^{(L)}$ (dashed). The code
is split into $B$ RGB-format triplets, processed by $B$ unmodified
RGB passes, then reassembled and decoded; $B{=}2$ at $k{=}6$.}
\label{fig:overview}
\end{figure*}

Fig.~\ref{fig:overview} summarises the system: a jointly trained
encoder--decoder pair around otherwise unmodified RGB render passes.
This section develops the constraints that shape it.

\subsection{Algebraic constraints from light transport}
\label{sec:algebra}

Path tracing repeatedly applies three operations to spectral
quantities: scaling (intensity, BSDF evaluation), addition (multiple
light contributions, path accumulation), and \emph{element-wise}
(Hadamard) multiplication: reflectance with incident radiance,
transmittance with through-flux, per-channel attenuation through
participating media, and conductor Fresnel evaluated at the complex
index $\eta(\lambda)\!-\!i\kappa(\lambda)$.\footnote{We use $\kappa$
for the imaginary part of the conductor index to avoid collision with
the latent dimension $k$.}
Discretising the visible spectrum on $n$ wavelength bins yields the
algebra $\mathcal{A}=(\R^n,+,\had)$ on which transport acts.

A renderer-friendly code must support the same three operations
\emph{directly on the compact codes}, so that the renderer can treat
codes as ordinary $k$-channel inputs. Formally, this requires the
encoder $\E$ to act as a ring homomorphism
$\mathcal{A}\to(\R^k,+,\had)$.

\subsection{Impossibility of exact algebra-preserving compression}
\label{sec:impossible}

The compression we seek does not exist exactly:

\begin{theorem}\label{thm:impossibility}
When $k<n$ there is no injective ring homomorphism
$\phi:(\R^n,+,\had)\to(\R^k,+,\had)$.
\end{theorem}

The argument is that the standard basis $\{e_1,\dots,e_n\}$ forms a
system of $n$ pairwise orthogonal nonzero idempotents under $\had$,
while $(\R^k,+,\had)$ admits at most $k$ such
idempotents~\cite{Lang2002,AtiyahMacdonald1969}, so for $k<n$ at
least one $e_i$ lies in $\ker\phi$. The full proof and algebraic
development are in Supplemental~Sec.~1.

The natural relaxation is \emph{distributional} preservation: scaling
and addition are preserved exactly (a property our linear architecture
below guarantees by construction), while Hadamard multiplication is
preserved approximately on the low-dimensional manifold of natural
visible spectra learned from data. The asymmetry is deliberate.
Scaling and addition enter every radiance estimate through path
accumulation and Monte-Carlo averaging, where a systematic error
would bias the estimator itself; the Hadamard product enters once
per material interaction, where a small approximation error is
perceptually tolerable, and the multi-bounce measurements in
\secref{sec:material_classes} show that it does not accumulate.

\subsection{Linear non-negative codec}
\label{sec:codec_arch}

We instantiate the encoder $\E$ and decoder $\D$ as linear maps
\begin{equation}
  z = \E(s) = W_\mathrm{enc}\,s, \qquad
  \hat s = \D(z) = W_\mathrm{dec}\,z,
  \label{eq:linear_codec}
\end{equation}
with $W_\mathrm{enc}\!\in\!\R^{k\times n}$ and
$W_\mathrm{dec}\!\in\!\R^{n\times k}$.
Linearity gives $\E(\alpha s)=\alpha\E(s)$ and
$\E(s_1\!+\!s_2)=\E(s_1)+\E(s_2)$ exactly; only the Hadamard property
is learned.

Both weight matrices are constrained non-negative through a softplus
reparameterisation of unconstrained learnable
matrices.\footnote{Parameterisation details, signed-decoder
ablations, and fidelity diagnostics are in
Supplemental~Sec.~2.}
Non-negativity is essential, not cosmetic: codes of non-negative
spectra remain non-negative, so they pack into RGB framebuffers and
flow through an unmodified forward-light-only renderer. The columns
of $W_\mathrm{dec}$ admit a direct interpretation as a basis of
spectra, with the latent code being the non-negative mixture-weight
vector for that basis.

\subsection{The codec as a diagonal reduced matrix}
\label{sec:fichet_view}

Fichet et al.~\shortcite{fichet2024fluorescence} describe spectral
reduction by a pair of maps: an analysis map (our $W_\mathrm{enc}$)
and a reconstruction basis (our $W_\mathrm{dec}$). In their
framework, a material whose spectral action is the diagonal operator
$\mathrm{diag}(m)$, with $m\in\R^n_{\ge 0}$ the material spectrum,
acts on the reduced coefficients through the $k\times k$
\emph{reduced matrix}
\begin{equation}
\begin{aligned}
  R_m &\;=\; W_\mathrm{enc}\,\mathrm{diag}(m)\,W_\mathrm{dec},\\
  (R_m)_{ij} &\;=\; \sum_\lambda
  W_\mathrm{enc}[i,\lambda]\; m(\lambda)\; W_\mathrm{dec}[\lambda,j].
\end{aligned}
\label{eq:reduced_matrix}
\end{equation}
Applying general reduced matrices requires shader-level
matrix--vector products, which is exactly the modification Fichet et
al.\ make. An unmodified RGB renderer can only multiply per channel,
so it applies $\mathrm{diag}(z_m)$: our setting is the restriction of
the reduced-matrix framework to \emph{diagonal} $R_m$.

Requiring $(R_m)_{ij}=0$ for all $i\neq j$ and for \emph{every}
non-negative material spectrum $m$ forces each summand in
Eq.~\eqref{eq:reduced_matrix} to vanish:
$W_\mathrm{enc}[i,\lambda]\,W_\mathrm{dec}[\lambda,j]=0$ at every
$\lambda$, i.e.\ encoder row $i$ and decoder column $j$ must have
disjoint supports for $i\neq j$. Non-negativity is what makes
this rigid: with signed matrices the off-diagonal integral could
cancel across wavelengths without the supports being disjoint,
whereas non-negative summands admit no cancellation. At the same
time, a faithful round-trip requires each encoder row to overlap its
\emph{own} decoder column. Exact diagonality for all $m$ therefore
forces a partition of the spectrum: each basis function owns a
wavelength range, which is precisely binning.

Binning, however, caps accuracy: the comparison in \secref{sec:capacity_choice} places every pure-binning variant at roughly $3\times$ our error. The codec relaxes exact diagonality to diagonality \emph{in distribution}: the training objective penalises off-diagonal leakage only on products of natural spectra, not for every conceivable $m$. This relaxation admits learned solutions that binning excludes, and the learning is where the accuracy is earned: in the same comparison, replacing the bin-average encoder with a learned one nearly halves the error of fixed boxes, and learning the decoder jointly nearly halves it again. The learned matrices keep nearly disjoint supports with small, deliberate overlaps (Supplemental~Sec.~2). The codec is best understood as the distributionally-diagonal member of the reduced-matrix family that minimises rendered colour error.

The codec is \emph{content-agnostic}: the same $\E,\D$ encode
reflectance $R$, illumination $L$, transmittance $\tau$, absorption
$\sigma_a$ and scattering $\sigma_s$ coefficients, and the spectral
parts of the conductor IOR $\eta,\kappa$ into a common latent space.
The learned matrices realise these constraints: the decoder columns
are smooth, band-localised spectra with the near-disjoint supports
derived above, spanning a chromaticity gamut larger than sRGB on the
green, blue and red edges, and the encoder rows are sparser and
multi-peaked, acting as band-pass analysers rather than as the
transpose of the decoder.

\subsection{Pass packing}
\label{sec:passpack}

We choose $k{=}3B$ so the latent code splits into $B$ blocks of three
channels. Each block is a triplet; the renderer treats it as an
ordinary RGB colour, applies its standard per-channel operations, and
writes the output to the corresponding block of an HDR framebuffer.
The full $k$-dimensional code is processed in $B$ unmodified RGB
passes. For two spectra $s_R,s_L$ encoded as $z_R,z_L\in\R^k$ the
renderer computes
\begin{equation}
  z_{R}\had z_{L}
  \;=\; \bigoplus_{b=1}^{B}\!\bigl(z_R^{(b)}\!\had\,z_L^{(b)}\bigr),
  \label{eq:passpack}
\end{equation}
with $z^{(b)}\!\in\!\R^3$ the $b$-th triplet and $\bigoplus$
concatenation.
Eq.~\eqref{eq:passpack} introduces no new operation: it is the
same $\had$ on a $k$-vector, written to expose the renderer's
per-pass three-channel structure.
The training objective, defined next, drives the codec toward
\begin{equation}
  \E(s_R\had s_L) \;\approx\; \E(s_R)\had\E(s_L),
  \label{eq:approx_hadamard}
\end{equation}
on the data manifold, so the per-pass product the renderer computes
matches the encoded product of the spectral interaction.

\subsection{Training objectives}
\label{sec:losses}

The codec is trained on a corpus of measured and synthetic reflectance
spectra (Munsell extended to sRGB-edge coverage) and an augmented
illumination corpus (LSPDD-measured lamps~\cite{LSPDD2026}, CIE
daylight, and synthetic narrowband variants, prefiltered by bin
integration so that mercury lines and other narrow peaks survive
downsampling). Full corpus
construction, split methodology, and representative test spectra are
in Supplemental~Sec.~3.3.

The codec minimises
\begin{equation}
  \mathcal{L} =
   \lambda_\mathrm{e2e}\,\mathcal{L}_\mathrm{e2e}
  +\lambda_\mathrm{rec}\,\mathcal{L}_\mathrm{rec}
  +\lambda_\mathrm{code}\,\mathcal{L}_\mathrm{code}
  +\lambda_\mathrm{col}\,\mathcal{L}_\mathrm{col},
  \label{eq:total_loss}
\end{equation}
each term targeting one component of the spectral pipeline:
\begin{align*}
  \mathcal{L}_\mathrm{e2e}  &= \Phi\bigl(\D(\E(R)\had\E(L)),\; R\had L\bigr)\\
  \mathcal{L}_\mathrm{rec}  &= \mathrm{MSE}(\D\E R,\,R)+\mathrm{MSE}(\D\E L,\,L)\\
  \mathcal{L}_\mathrm{code} &= \mathrm{MSE}\bigl(\E(R)\had\E(L),\;\E(R\had L)\bigr)\\
  \mathcal{L}_\mathrm{col}  &= \Delta E_{76}\bigl(\D(\E(R)\had\E(L)),\; R\had L\bigr).
\end{align*}
$\mathcal{L}_\mathrm{e2e}$ is the end-to-end fit between the
codec-rendered product and the spectral ground truth, with
$\Phi(x,y)=(2-\cos(x,y))\,\mathrm{MSE}(x,y)$ combining magnitude and
shape. $\mathcal{L}_\mathrm{rec}$ enforces round-trip accuracy on
individual spectra. $\mathcal{L}_\mathrm{code}$ is the
Hadamard-multiplicativity term that Theorem~\ref{thm:impossibility}
forbids from reaching zero in general but that we minimise in
distribution. $\mathcal{L}_\mathrm{col}$ evaluates the same
product--ground-truth pair as $\mathcal{L}_\mathrm{e2e}$, but
perceptually: as $\Delta E_{76}$ in CIELAB under the per-sample
illuminant white point.
$\mathcal{L}_\mathrm{code}$ alone admits the trivial collapse
$\E\equiv 0$; $\mathcal{L}_\mathrm{e2e}$ and $\mathcal{L}_\mathrm{rec}$
make this infeasible by requiring the codec to reproduce both
individual spectra and their products. The same argument pins the
overall scale: rescaling $\E\!\to\!s\E$ and $\D\!\to\!\D/s$ leaves
$\mathcal{L}_\mathrm{rec}$ unchanged, but
$\mathcal{L}_\mathrm{e2e}$ and $\mathcal{L}_\mathrm{col}$ apply the
encoder twice and the decoder once, so their error grows as $s$
shrinks and the degeneracy is excluded. Non-negativity is enforced
architecturally by the softplus parameterisation and needs no penalty.

Eq.~\eqref{eq:total_loss} trains on single reflectance--illuminant
products; it is $\mathcal{L}_\mathrm{code}$ that trains the latent
product structure which iterates over multiple bounces, and
multi-bounce accuracy is measured empirically (\secref{sec:material_classes}). The corpus contains reflectance
and illumination spectra only; transmittance, complex IOR, and
scattering coefficients are encoded with the same matrices, without
retraining (\secref{sec:materials}).

The four weights $\lambda_\bullet$ were selected by a two-phase random search over
130 codec variants, optimising for stability under chained products
(three or more spectra, i.e.\ multiple rendering bounces); the
search ranked candidates by $\Delta E_{94}$ for speed, while all
reported results use $\Delta E_{2000}$. The selected values, the
search protocol, and the multi-bounce
sensitivity sweep are in Supplemental~Sec.~3.2.

%% file: sections/material_encoding.inc
\section{Material-Class Encoding}
\label{sec:materials}

The codec compresses any visible-range spectral quantity the same way
(\secref{sec:algebra}): whichever spectral input a BSDF or volume model
expects (reflectance, transmittance, complex index of refraction,
absorption or scattering coefficients) is encoded once into a
$k$-dimensional latent code and processed by the unmodified renderer
in $B{=}k/3$ standard RGB passes. The renderer's per-channel arithmetic
(scaling, addition, Hadamard products at material interactions, and
per-channel attenuation through participating media) operates on the
$k$ latent channels exactly as it would on three RGB channels.

\paragraph{Coloured glass.}
\label{sec:materials:glass}
Coloured glass is characterised by a wavelength-dependent transmittance
$\tau(\lambda)\!\in\![0,1]$ that attenuates light passing through it.
We handle two physically distinct regimes. For thin filters and
surface coatings, the transmittance acts as a per-crossing multiplier
on the radiance: at each interface
$L_\mathrm{out}(\lambda)\!=\!\tau(\lambda)\,L_\mathrm{in}(\lambda)$.
Encoding $z_\tau\!=\!\E(\tau)$ once and applying $z_\tau^{(b)}$
per channel across $N$ surface crossings corresponds to integer
Hadamard powers $z_\tau^{N}$, the iterated-product structure the
codec is trained to approximate. For bulk-absorbing glass, spectral
attenuation follows the Beer--Lambert law along the optical path:
$T(d,\lambda)\!=\!\exp(-\sigma_a(\lambda)\,d)$. We obtain a latent
absorption coefficient by inverting the encoded transmittance at a
reference depth,
$\sigma_a^{(b)}\!=\!-\ln(z_\tau^{(b)})/d_\mathrm{ref}$; the renderer's
per-channel volume integrator then applies $\exp(-\sigma_a^{(b)}\,d)$
along each ray segment of length $d$, a fractional Hadamard power
the codec approximates.

\paragraph{Conductors.}
\label{sec:materials:conductor}
A conductor BSDF is parameterised by a wavelength-dependent complex
index of refraction $\eta(\lambda)\!-\!i\kappa(\lambda)$. We encode
$\eta$ and $\kappa$ \emph{as ordinary spectra} through the same codec,
recovering two $k$-dimensional codes $z_\eta,z_\kappa$, and feed them
per-pass to the unmodified conductor BSDF.
The Fresnel equations are non-linear in $\eta,\kappa$, but the
non-linearity acts \emph{per channel}: the renderer evaluates Fresnel
on the $k$ latent channels just as it would on three RGB
channels.\footnote{Per-channel Fresnel evaluation on RGB primaries is
known to distort metal colours~\cite{Hoffman2019Fresnel}; the
encoding applies equally when the renderer uses the Schlick
approximation.}
Replacing one accurate spectral Fresnel evaluation with $k$ approximate
per-channel ones is far less lossy than the standard RGB collapse
because the $k$ near-disjoint latent channels sample $\eta$ and
$\kappa$ much more finely than three broad RGB primaries. The
renderer remains an opaque RGB BSDF evaluator; the spectral
information lives in the latent channels alone.

\paragraph{Homogeneous participating media.}
\label{sec:materials:sss}
Absorption $\sigma_a(\lambda)$ and scattering $\sigma_s(\lambda)$
coefficients are ordinary non-negative visible-range spectra, which the codec embeds directly. We encode each separately, and the renderer's standard
per-channel volume integrator (the same code path it uses for an
RGB homogeneous medium) accumulates per-channel transmittance
$\exp(-(\sigma_a^{(b)}\!+\!\sigma_s^{(b)})\,d)$ and applies the
single-scatter albedo
$\sigma_s^{(b)}/(\sigma_a^{(b)}\!+\!\sigma_s^{(b)})$ as a per-channel
multiplier at scattering events, the transport that produces
subsurface scattering (SSS) in translucent materials. Media
therefore exercise two
per-channel regimes: extinction follows the Beer--Lambert analysis
above, and the scattering albedo is a per-channel ratio of two
encoded spectra. The same codec, trained on reflectance and
illumination only, encodes extinction coefficients without
retraining. For media with strongly saturated absorption bands,
whose albedo the ratio regime cannot represent accurately, we
factor the transmittance into a per-crossing chromatic term, encoded
as a surface transmittance, and a smooth residual absorption
coefficient carried by the medium.

\paragraph{Heterogeneous participating media.}
\label{sec:materials:hetero}
Per-voxel spectral data is encoded into the same $k$-dimensional
latent space and stored as $B$ RGB-format grid volumes; each pass
reads its triplet through the renderer's heterogeneous-medium
machinery without modification. Linearity of the encoder makes this
both practical and exact. Voxel spectra are typically blends of a
few feature spectra, and encoding commutes with blending: the
feature spectra are encoded once and their latent codes blended per
voxel, so no per-voxel encoding is needed, and the renderer's
volume filtering commutes with the encoding for the same reason
(exactly so for trilinear and any non-negative linear filter). Along a ray, the accumulated optical depth is a per-channel
sum of voxel contributions, and sums are preserved exactly, as is
rescaling by a scalar density field, which multiplies all channels
by the same factor. Heterogeneity
therefore introduces no approximation beyond the homogeneous
regimes above: the deviation remains that of the per-channel
exponential and the albedo ratio.

Per-class encoding parameters and renderer bindings are documented
in Supplemental~Sec.~6.

%% file: sections/upsampling.inc
\section{Latent Upsampling for Legacy RGB Assets}
\label{sec:upsampling}

While our codec compresses spectral inputs end-to-end, production
pipelines hold large libraries of RGB textures, materials, and
environment maps for which spectral data is unavailable.
The obvious route for such assets is to uplift them with an existing
RGB-to-spectrum method, for example Jakob and
Hanika~\shortcite{Jakob2019spectral}, and encode the result. This
composes two independent approximations: the uplift selects one
metamer with no regard for the codec, and the encoder then
compresses a spectrum that need not lie near its training
distribution. We instead train lightweight networks that map the
RGB input directly to a latent code. The prediction target is the
code the codec itself would produce for the underlying spectrum, so
the networks learn to land on codes that render accurately: on
held-out data the composed route yields about $1.6\times$ the
colour error of direct prediction (the through-codec baseline rows of
Tab.~\ref{tab:upsampling_dE}), while direct prediction stays close
to the codec's own accuracy on ground-truth spectra. The upsamplers
are a convenience bridge for RGB-only assets, not a spectral
reconstruction method.

Both networks map a colour input
$\mathbf{c}\in\R^3$ to a $k$-dimensional latent code:
\begin{equation}
  z_\mathrm{pred} =
  \begin{cases}
    f^{(\mathrm{R})}_\theta(\mathbf{c}) & \text{(reflectance assets)}\\
    f^{(\mathrm{L})}_\phi(\mathbf{c})   & \text{(illumination assets).}
  \end{cases}
  \label{eq:upsamplers}
\end{equation}
The networks operate in linear XYZ: at training time the input is
the CMF projection of the training spectrum, and at render time RGB
assets are first linearised and converted with the standard
sRGB-to-XYZ matrix. We use two networks rather than one because reflectance
distributions are bounded ($s\in[0,1]^n$) and broadly Munsell-like,
whereas illumination is unbounded and includes narrowband emission
peaks; the two need different inductive biases.
Both networks are per-pixel multilayer perceptrons with scale
equivariance (the input magnitude is factored out, the network maps the
unit direction to a unit-scale latent, and the magnitude is reapplied at
the output) so the codec's exact scale equivariance carries through to
RGB inputs.
The reflectance network additionally has an \emph{auxiliary spectrum
head}, a small MLP that decodes the predicted latent back to a
reconstructed spectrum, supervised by a smoothness penalty
during training.
This auxiliary path acts as a regulariser tied to the smooth Munsell-like
prior of reflectance and improves reflectance accuracy by $2.5\times$
(Supplemental~Sec.~7); we find no equivalent benefit for
illumination, whose narrowband peaks penalise smoothness, so
$f^{(\mathrm{L})}$ uses a single latent head.
Codec weights are frozen during upsampler training.

\paragraph{Training objective.}
For each spectrum $s$ in the codec's training set the input is the
CMF projection $\mathbf{c}=T_\mathrm{cmf}\,s$ and the target latent
is $z_\mathrm{gt}=\E(s)$.
Both networks share a common latent-supervision loss
\begin{equation}
\label{eq:upsample_loss}
\begin{aligned}
\mathcal{L}_\mathrm{lat} ={}&
  \mathrm{MSE}\bigl(z_\mathrm{pred},z_\mathrm{gt}\bigr) \\
  &+ \lambda_\mathrm{maxabs}\,\mathbb{E}\bigl[\|z_\mathrm{pred}-z_\mathrm{gt}\|_\infty\bigr] \\
  &+ \lambda_\mathrm{col}\,\Delta E_{76}\bigl(\D(z_\mathrm{pred}),\D(z_\mathrm{gt})\bigr),
\end{aligned}
\end{equation}
where the MSE enforces overall latent accuracy, the $\ell_\infty$ term
controls outlier dimensions, and the perceptual $\Delta E_{76}$ term,
computed on the decoded spectra in CIELAB, ensures colorimetric
stability after decode.
The reflectance network additionally supervises its spectrum-head
output $\hat{s}_\mathrm{pred}$ against the ground-truth spectrum:
$\mathcal{L}_\mathrm{spec}^{(\mathrm{R})} = \mathrm{MSE}(\hat{s}_\mathrm{pred}, s) + \lambda_\mathrm{smooth}\,\sum_i (\hat{s}_{i+1}-\hat{s}_i)^2$,
yielding the total reflectance loss
$\mathcal{L}^{(\mathrm{R})} = \mathcal{L}_\mathrm{lat} + \lambda_\mathrm{spec}\,\mathcal{L}_\mathrm{spec}^{(\mathrm{R})}$.
Loss weights and training schedule are in Supplemental~Sec.~7.

\paragraph{Scope.}
Three boundaries delimit what these networks do. First,
RGB-to-spectrum upsampling is one-to-many: a colour corresponds to
many metamers, and a deterministic network selects a single one.
Like all one-to-one upsamplers, ours does not address metamerism;
when an asset's true spectrum differs from the selected metamer,
relighting can diverge, and smooth selections are the safer choice.
We therefore claim no recovery of true spectra: the target is a
latent code that reproduces the asset's colour and renders robustly
across illuminants. Second, unlike methods that enforce the input
colour as a hard constraint, our round-trip is soft:
Eq.~\eqref{eq:upsample_loss} penalises but does not eliminate colour
drift. The cost is small on reflectances (Supplemental~Sec.~7) and is most visible
under broadband illumination upsampling, where a hard-constrained
uplift is close to exact by construction
(Tab.~\ref{tab:upsampling_dE}); in exchange, the soft formulation
is free to trade exact colour reproduction at one condition for
rendering accuracy across conditions. Third, scale equivariance is the correct prior for
illumination, whose overall intensity is arbitrary, but for
reflectance it forgoes the brightness information that recent
uplift methods exploit to constrain the feasible metamer set; we
accept this trade for a single shared mechanism across both asset
types.

\paragraph{Application: spectral assets from RGB inputs.}
At render time, RGB textures and environment maps are upsampled per
pixel into a $k$-dimensional latent map, packed as $k/3$ RGB-format EXR
textures, and rendered through the unmodified renderer
(\secref{sec:results_upsampling}).

%% file: sections/experiments.inc
\section{Experiments}
\label{sec:experiments}

We evaluate the codec against spectral rendering as ground truth (Mitsuba\,0.6 spectral build) and against standard RGB rendering as the practical baseline (the same Mitsuba\,0.6 in RGB mode; the
spectral build is never touched by our pipeline). We use $k{=}6$ (two RGB passes) throughout. All qualitative figures decode latents to spectra and CMF-integrate to sRGB. Sample budgets vary
per scene class to balance Monte-Carlo noise against render time and are stated per figure; further tables and ablations are in the supplemental material.

\newlength{\gcellW}\setlength{\gcellW}{0.163\textwidth}
\newlength{\imgcellH}\setlength{\imgcellH}{0.1222\textwidth}
\newlength{\glabW}\setlength{\glabW}{0.024\textwidth}
\newlength{\crlh}\setlength{\crlh}{0.135\textwidth}
\newlength{\handcellW}\setlength{\handcellW}{0.49\columnwidth}
\newlength{\cboxh}\setlength{\cboxh}{0.285\columnwidth}
\newlength{\cboxhi}\setlength{\cboxhi}{0.3407\columnwidth}  
\newlength{\cboxp}\setlength{\cboxp}{0.278\columnwidth}   

\subsection{Codec reconstruction quality}
\label{sec:results_codec}

\begin{figure}[t]
  \centering
    \setlength{\tabcolsep}{1pt}
    \renewcommand{\arraystretch}{0}
    \begin{tabular}{@{}c@{\,}c@{\,}c@{\,}c@{}}
       \scriptsize Spectral GT &
       \scriptsize patches &
       \scriptsize Ours, $k{=}6$ &
       \scriptsize $\Delta E_{2000}$ \\[1pt]
       \includegraphics[height=\cboxp]{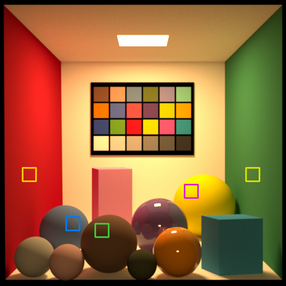} &
       \includegraphics[height=\cboxp]{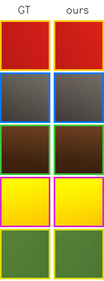} &
       \includegraphics[height=\cboxp]{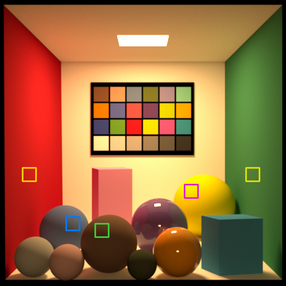} &
       \includegraphics[height=\cboxp]{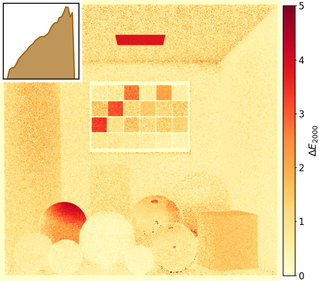} \\
       \includegraphics[height=\cboxp]{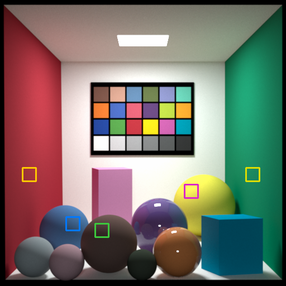} &
       \includegraphics[height=\cboxp]{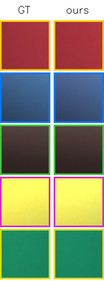} &
       \includegraphics[height=\cboxp]{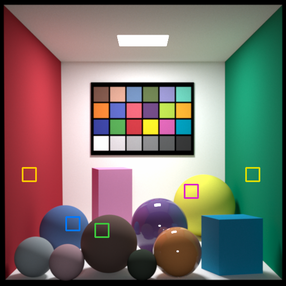} &
       \includegraphics[height=\cboxp]{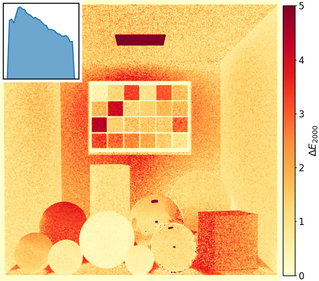} \\
       \includegraphics[height=\cboxp]{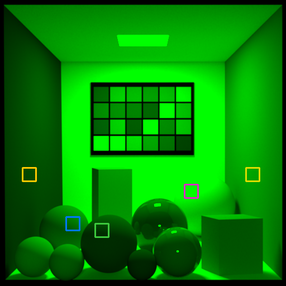} &
       \includegraphics[height=\cboxp]{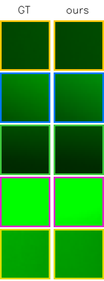} &
       \includegraphics[height=\cboxp]{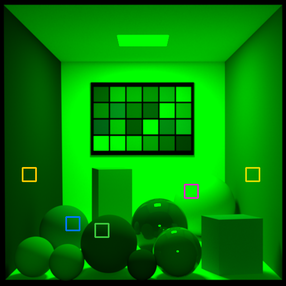} &
       \includegraphics[height=\cboxp]{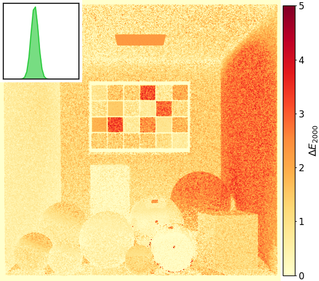} \\
       \includegraphics[height=\cboxp]{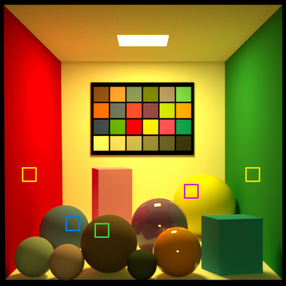} &
       \includegraphics[height=\cboxp]{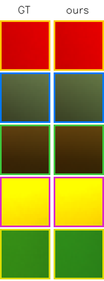} &
       \includegraphics[height=\cboxp]{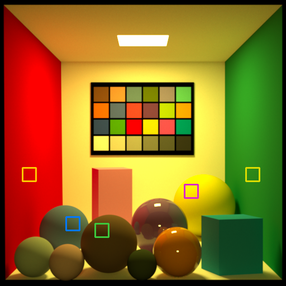} &
       \includegraphics[height=\cboxp]{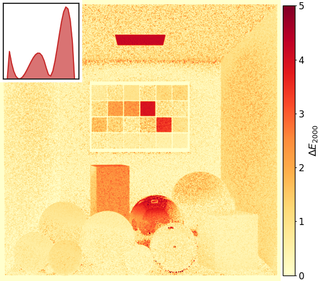} \\
       \includegraphics[height=\cboxp]{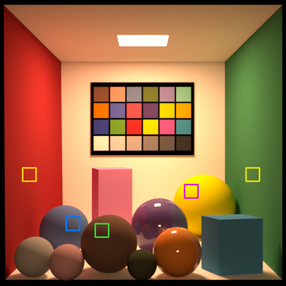} &
       \includegraphics[height=\cboxp]{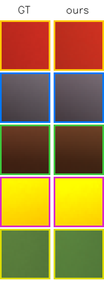} &
       \includegraphics[height=\cboxp]{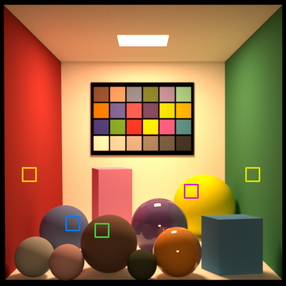} &
       \includegraphics[height=\cboxp]{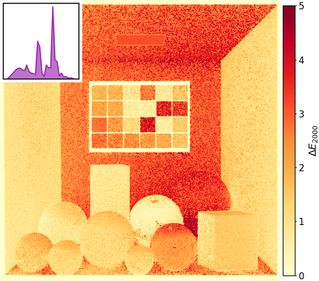} \\
    \end{tabular}
    \caption{\textbf{Cornell box under five illuminants.} Top to
    bottom: daylight 3356\,K, daylight 7150\,K, narrowband Gaussian
    at 525\,nm (FWHM 40\,nm), the structured red-dominant source
    narrow$_{351}$ (main peak 670\,nm), and CIE FL11. The middle
    column shows five magnified patches (marked in the renderings;
    top to bottom: red wall, blue sphere, shadowed sphere, yellow
    sphere, green wall) as adjacent ground-truth/ours pairs for
    direct colour comparison. Each $\Delta E_{2000}$ map embeds the
    illuminant SPD as an inset (top-left). Per-illuminant means and
    the RGB baseline are given in Tab.~\ref{tab:binning_main}.
    1024\,spp.}
    \label{fig:cbox_calibration}
\end{figure}

\paragraph{Cornell-box evaluation under five illuminants.}
We first evaluate the codec at $k{=}6$ on a Cornell box containing diverse BRDFs and a ColorChecker chart, rendered under broadband, narrowband, structured, and multi-peaked illumination (Fig.~\ref{fig:cbox_calibration}; per-condition values in the figure). Materials and illuminants here are spectral ground truth encoded directly through the codec, with no upsampling involved. Across all five illuminants the latent pipeline reduces $\overline{\Delta E}_{2000}$ vs.\ spectral ground truth by $1.6$--$4.4\times$ over standard RGB rendering; the advantage is largest under the structured red-dominant source narrow$_{351}$ ($0.86$ vs.\ $3.82$) and warm daylight ($0.90$ vs.\ $3.40$). A 52-cell sweep over Gaussian-narrowband illuminants (13 centres $\times$ 4 FWHMs spanning the visible band) gives a mean per-cell codec advantage of $11.2\times$ over RGB (Supplemental~Sec.~5).

\begin{figure}[t]
  \centering
  \setlength{\tabcolsep}{1.5pt}
  \begin{tabular}{@{}ccccc@{}}
    \includegraphics[width=0.19\columnwidth]{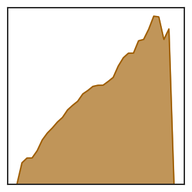} &
    \includegraphics[width=0.19\columnwidth]{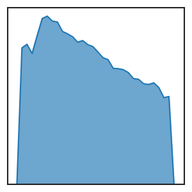} &
    \includegraphics[width=0.19\columnwidth]{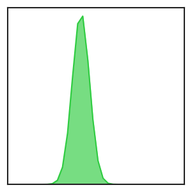} &
    \includegraphics[width=0.19\columnwidth]{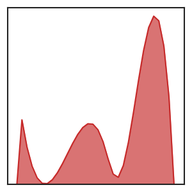} &
    \includegraphics[width=0.19\columnwidth]{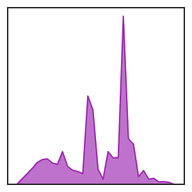} \\
    \scriptsize broad$_{12}$ &
    \scriptsize broad$_{63}$ &
    \scriptsize narrow$_{525}$ &
    \scriptsize narrow$_{351}$ &
    \scriptsize FL11 \\
  \end{tabular}
  \caption{\textbf{The five Cornell-box evaluation illuminants of
  Tab.~\ref{tab:binning_main} and
  Fig.~\ref{fig:cbox_calibration}}, normalised SPDs over
  380--730\,nm: a warm broadband (3356\,K), a daylight broadband
  (7150\,K), a 525\,nm Gaussian (FWHM 40\,nm), the structured red-dominant source narrow$_{351}$ (main peak 670\,nm, secondary lobes near 540 and 410\,nm),
  and the multi-peaked fluorescent FL11.}
  \label{fig:illum_spds}
\end{figure}

\paragraph{Capacity and binning baselines.}
\label{sec:capacity_choice}
Since the learned decoder has nearly disjoint supports
(\secref{sec:fichet_view}), a natural question is whether plain
wavelength binning at $k{=}6$ would suffice. It does not.
Tab.~\ref{tab:binning_main} evaluates the full ladder on the same
Cornell-box suite (illuminant SPDs in Fig.~\ref{fig:illum_spds}):
three pure
binning variants (bin-average encoder, box decoder, no training),
the two half-learned bases (fixed decoder shape, trained encoder),
and the jointly learned codec.

\begin{table}[t]
\centering
\caption{\textbf{Cornell-box mean $\Delta E_{2000}$ per illuminant for
$k{=}6$ representations.} Pure binning uses a bin-average encoder and
box decoder with no training; the adaptive variant optimises the bin
edges on the training illuminants; the ``bins at decoder transitions'' variant places bin
edges at our learned decoder's own transition wavelengths. All rows share the pipeline and ground truth of
Fig.~\ref{fig:cbox_calibration}, at 1024\,spp.}
\label{tab:binning_main}
\small
\resizebox{\columnwidth}{!}{%
\begin{tabular}{lrrrrrr}
\toprule
$k{=}6$ representation & broad$_{12}$ & broad$_{63}$ & narrow$_{525}$ & narrow$_{351}$ & FL11 & avg \\
\midrule
equal-width bins        & 2.02 & 2.18 & 1.73 & 10.47 & 5.36 & 4.35 \\
bins at decoder transitions & 1.86 & 1.98 & 2.92 & 9.23 & 5.00 & 4.20 \\
adaptive-width bins     & 1.98 & 2.07 & 1.33 & 6.69 & 5.57 & 3.53 \\
RGB baseline ($k{=}3$)  & 3.40 & 2.42 & 2.89 & 3.82 & 4.01 & 3.31 \\
equal boxes, learned encoder & 1.25 & 2.03 & 2.94 & 1.42 & 4.14 & 2.36 \\
Gaussian basis, learned encoder & 0.92 & 1.81 & 1.50 & 1.10 & 7.52 & 2.57 \\
ours, jointly learned   & \textbf{0.90} & \textbf{1.52} & \textbf{1.11} & \textbf{0.86} & \textbf{1.83} & $\mathbf{1.24}$ \\
\bottomrule
\end{tabular}}
\end{table}

\begin{figure*}[t]
  \centering
  \captionsetup[subfigure]{position=top,skip=2pt}
  \begin{subfigure}[t]{0.135\textwidth}
    \centering
    \subcaption{Spectral GT\\reference}
    \includegraphics[width=\linewidth]{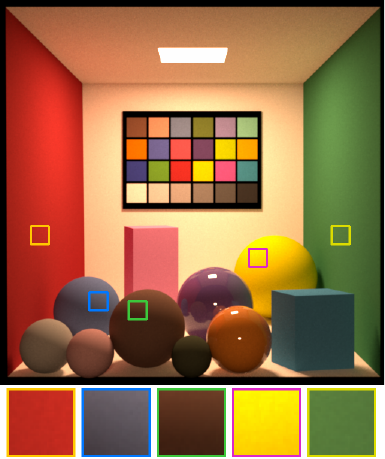}
  \end{subfigure}\hfill
  \begin{subfigure}[t]{0.135\textwidth}
    \centering
    \subcaption{Ours ($k{=}6$)\\jointly learned}
    \includegraphics[width=\linewidth]{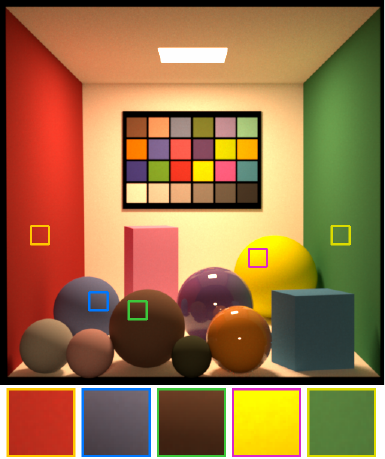}
  \end{subfigure}\hfill
  \begin{subfigure}[t]{0.135\textwidth}
    \centering
    \subcaption{Equal boxes\\learned encoder}
    \includegraphics[width=\linewidth]{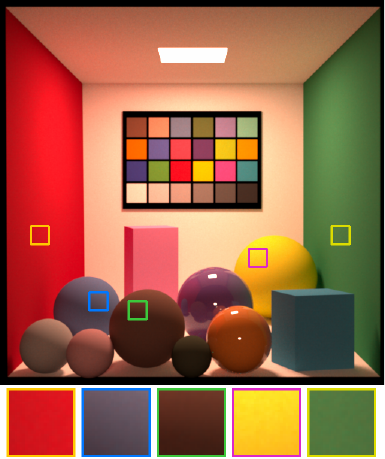}
  \end{subfigure}\hfill
  \begin{subfigure}[t]{0.135\textwidth}
    \centering
    \subcaption{Gaussian basis\\learned encoder}
    \includegraphics[width=\linewidth]{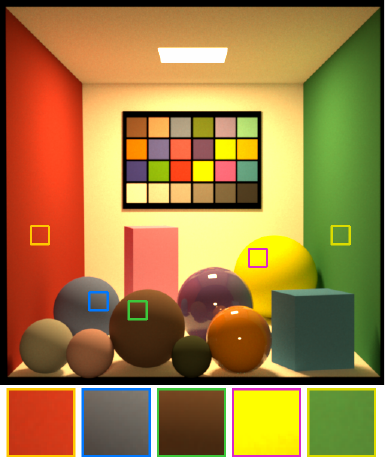}
  \end{subfigure}\hfill
  \begin{subfigure}[t]{0.135\textwidth}
    \centering
    \subcaption{Equal boxes\\untrained encoder}
    \includegraphics[width=\linewidth]{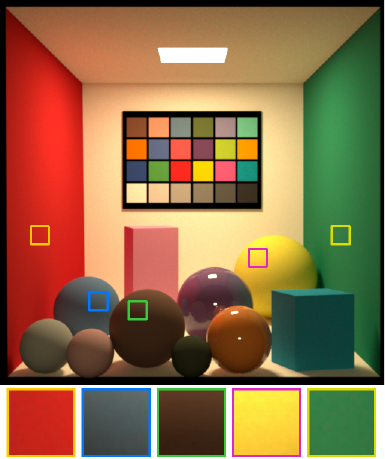}
  \end{subfigure}\hfill
  \begin{subfigure}[t]{0.135\textwidth}
    \centering
    \subcaption{Gaussian basis\\untrained encoder}
    \includegraphics[width=\linewidth]{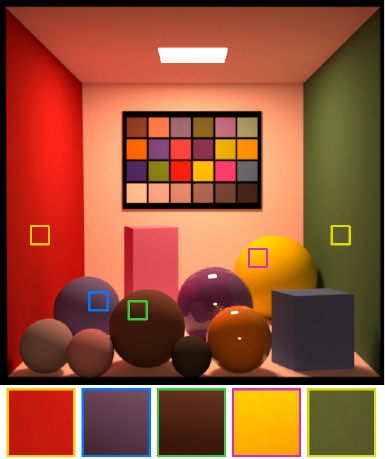}
  \end{subfigure}\hfill
  \begin{subfigure}[t]{0.135\textwidth}
    \centering
    \subcaption{RGB baseline\\($k{=}3$)}
    \includegraphics[width=\linewidth]{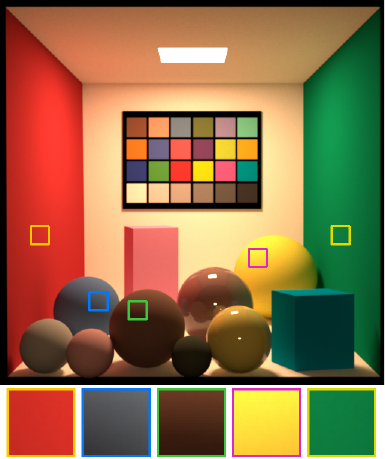}
  \end{subfigure}
  \\[0.4em]
  \begin{subfigure}[t]{0.135\textwidth}
    \centering
    \includegraphics[width=\linewidth]{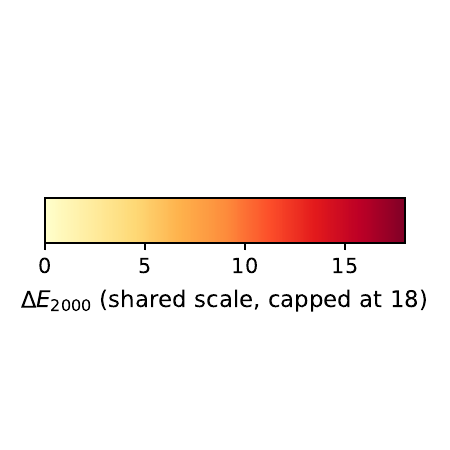}
  \end{subfigure}\hfill
  \begin{subfigure}[t]{0.135\textwidth}
    \centering
    \includegraphics[width=\linewidth]{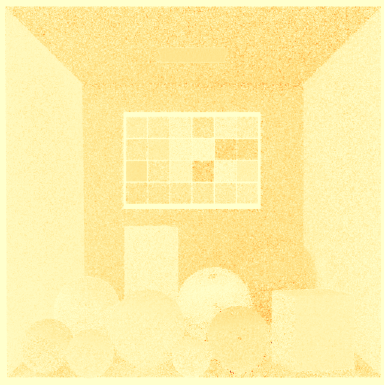}
  \end{subfigure}\hfill
  \begin{subfigure}[t]{0.135\textwidth}
    \centering
    \includegraphics[width=\linewidth]{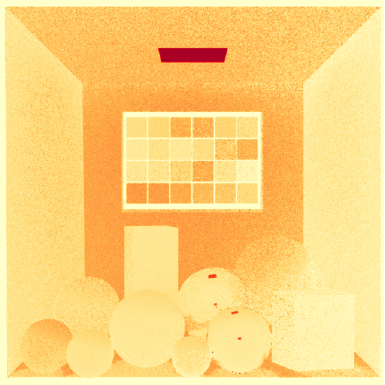}
  \end{subfigure}\hfill
  \begin{subfigure}[t]{0.135\textwidth}
    \centering
    \includegraphics[width=\linewidth]{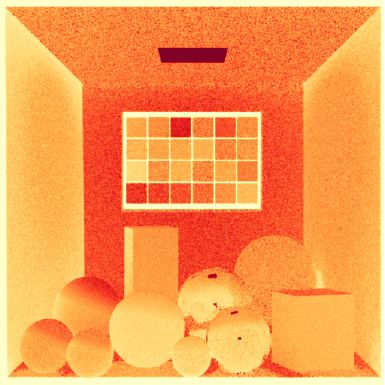}
  \end{subfigure}\hfill
  \begin{subfigure}[t]{0.135\textwidth}
    \centering
    \includegraphics[width=\linewidth]{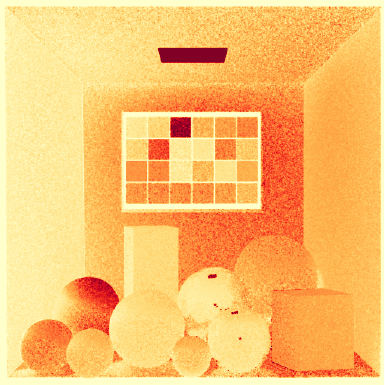}
  \end{subfigure}\hfill
  \begin{subfigure}[t]{0.135\textwidth}
    \centering
    \includegraphics[width=\linewidth]{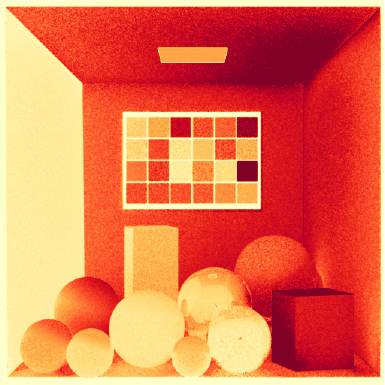}
  \end{subfigure}\hfill
  \begin{subfigure}[t]{0.135\textwidth}
    \centering
    \includegraphics[width=\linewidth]{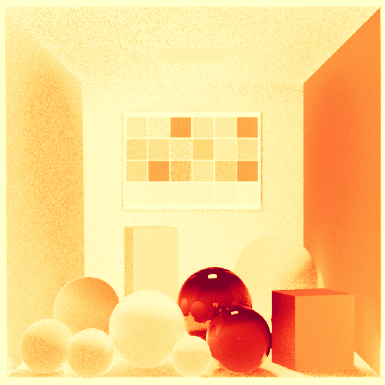}
  \end{subfigure}
  \caption{\textbf{Capacity baselines on the Cornell box under
  FL11.} Top row: the spectral reference (a), our codec (b), the two
  encoder-trained bases of Tab.~\ref{tab:binning_main} (c,\,d), the
  same bases with untrained encoders (e,\,f; Supplemental~Tab.~3), and the
  RGB baseline (g). Bottom row: each render's $\Delta E_{2000}$ map
  against (a); all maps share the colour scale shown under (a),
  capped at $\Delta E_{2000}{=}18$. The five magnified patches under
  each rendering (locations as in Fig.~\ref{fig:cbox_calibration})
  allow direct colour comparison; per-variant means are in
  Tab.~\ref{tab:binning_main} and Supplemental~Tab.~3, and the FL11 SPD is
  in Fig.~\ref{fig:illum_spds}.}
  \label{fig:capacity_renders}
\end{figure*}

Averaged over the five illuminants, every binning variant, including
bins placed at our own decoder's transitions, is about $3\times$
less accurate than the learned codec and worse than plain RGB. The
two narrowband columns expose why: binning accuracy depends on how
the source's spectral structure falls relative to the fixed bin
edges. The same adaptive bins score $1.33$ when the $525$\,nm
Gaussian lands near a bin centre and $6.69$ on narrow$_{351}$, whose
red-dominant multi-lobed spectrum straddles several bin edges, and
no placement resolves FL11's three emission peaks ($5.0$--$5.6$
vs.\ our $1.8$). The codec is placement-robust ($1.11$ and $0.86$)
and remains the most accurate representation even in the
binning-friendly $525$\,nm column. Nor is the decoder shape the
whole contribution: on the same equal-width boxes, learning only the
encoder cuts the average error nearly in half ($4.35\!\to\!2.36$),
and jointly learning the decoder nearly halves it again
($\to\!1.24$). The half-learned bases also expose a robustness
failure the joint codec avoids: each is competitive on smooth
broadband light but collapses on a different structured cell (boxes
at $2.94$ on the $525$\,nm source and $4.14$ on FL11, Gaussians at
$7.52$ on FL11). The near-disjoint decoder is an outcome of the
joint optimisation, not evidence that a box basis suffices.
Fig.~\ref{fig:capacity_renders} shows the corresponding renders and
error maps under FL11, the column where the differences are largest.
PCA and NMF baselines and
further per-illuminant breakdowns are in Supplemental~Sec.~4.

\paragraph{Capacity at $k{=}6$ vs.\ $k{=}9$.}
A 13-scene cross-class sweep (the four Cornell-box
illuminants, three coloured glasses, three measured conductors,
and three subsurface materials) gives an aggregate
$\overline{\Delta E}_{2000}$ difference of $-0.05$ in favour of
$k{=}9$; the gain is negligible, and per-scene differences trade in
both directions rather than favouring either setting systematically
(Supplemental~Sec.~4.4, which also examines why the gain saturates).
At $50\%$ extra rendering cost this is not a favourable trade; we
use $k{=}6$ throughout.

\subsection{Material-class evaluation}
\label{sec:material_classes}

We evaluate the codec on the material classes of \secref{sec:materials}
with one scene per material, the XYZ RGB Inc.\ dragon (the
heterogeneous SSS case uses a hand volume): coloured glass in the
surface and Beer--Lambert regimes, subsurface
scattering in the dipole and homogeneous regimes, measured
conductors, and a heterogeneous volume. All scenes use the same
$k{=}6$ codec under D65; the reference is Mitsuba\,0.6's spectral
build. We report the object-masked mean $\Delta E_{2000}$ against
the reference after matching each candidate's mean luminance to it
on the mask; the remaining materials of each class are in
Supplemental~Sec.~6.4.

\begin{figure*}[t!]
  \centering
  \setlength{\tabcolsep}{1pt}
  \renewcommand{\arraystretch}{1}
  \begin{tabular}{@{}c@{\,}c@{\,}c@{\,}c@{\,}c@{\,}c@{\,}c@{\,}c@{}}
    & \scriptsize Spectral GT
    & \scriptsize Ours (latent, $k{=}6$)
    & \scriptsize RGB baseline
    & \scriptsize patches
    & \scriptsize $\Delta E_{2000}$ Latent
    & \scriptsize $\Delta E_{2000}$ RGB & \\[1pt]
    \parbox[b][\imgcellH][c]{\glabW}{\centering\rotatebox{90}{\small OG550 B--L}} &
    \includegraphics[width=\gcellW]{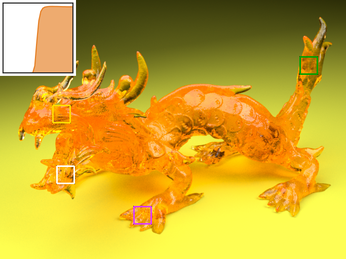} &
    \includegraphics[width=\gcellW]{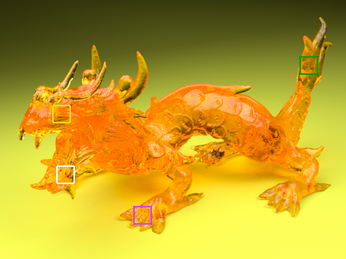} &
    \includegraphics[width=\gcellW]{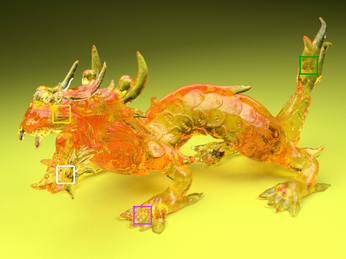} &
    \includegraphics[height=\imgcellH]{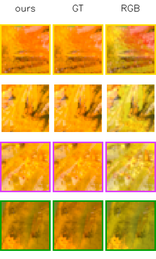} &
    \includegraphics[width=\gcellW]{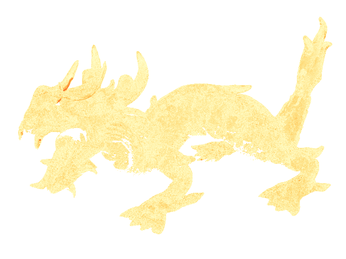} &
    \includegraphics[width=\gcellW]{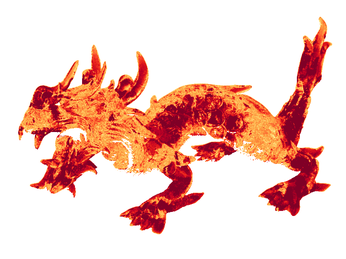} &
    \includegraphics[height=\imgcellH]{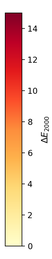} \\[1pt]
    \parbox[b][\imgcellH][c]{\glabW}{\centering\rotatebox{90}{\small OG550 surface}} &
    \includegraphics[width=\gcellW]{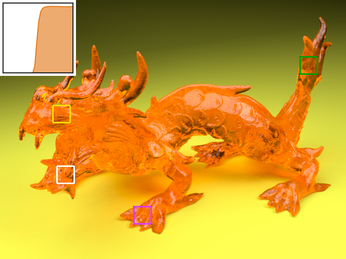} &
    \includegraphics[width=\gcellW]{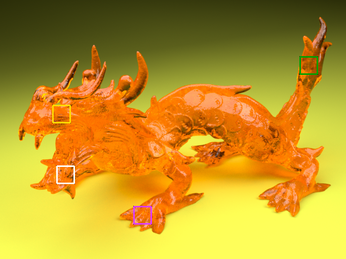} &
    \includegraphics[width=\gcellW]{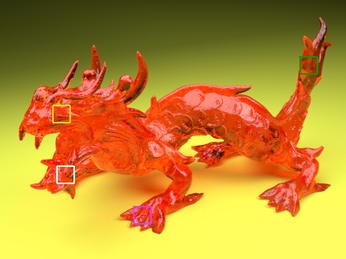} &
    \includegraphics[height=\imgcellH]{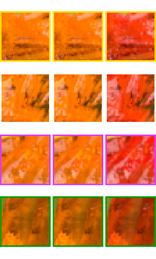} &
    \includegraphics[width=\gcellW]{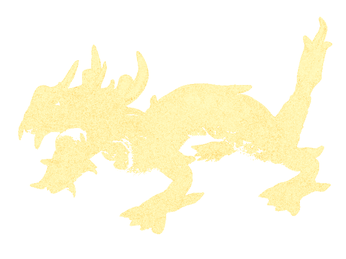} &
    \includegraphics[width=\gcellW]{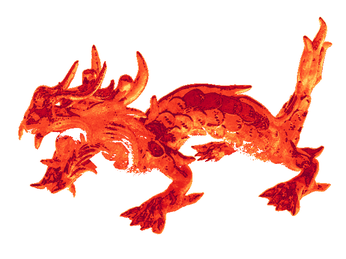} &
    \includegraphics[height=\imgcellH]{figures/material_classes/colorbar_main.png} \\[1pt]
    \parbox[b][\imgcellH][c]{\glabW}{\centering\rotatebox{90}{\small BG7 B--L}} &
    \includegraphics[width=\gcellW]{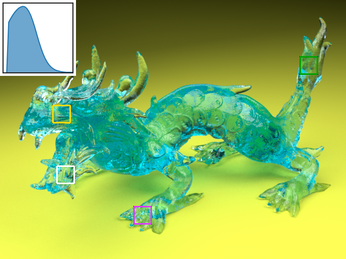} &
    \includegraphics[width=\gcellW]{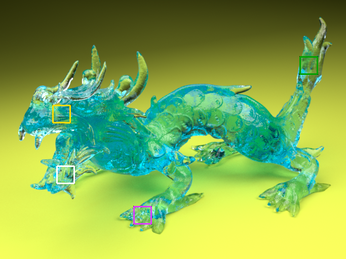} &
    \includegraphics[width=\gcellW]{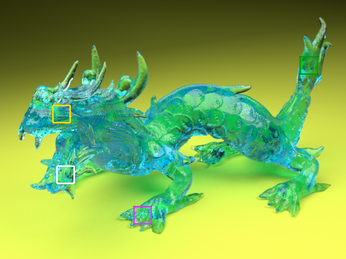} &
    \includegraphics[height=\imgcellH]{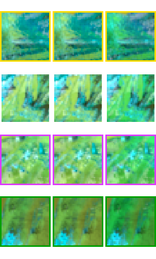} &
    \includegraphics[width=\gcellW]{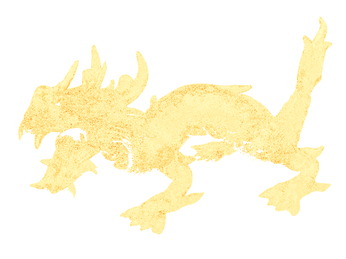} &
    \includegraphics[width=\gcellW]{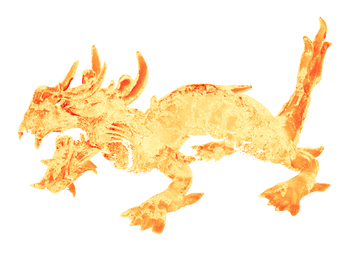} &
    \includegraphics[height=\imgcellH]{figures/material_classes/colorbar_main.png} \\[1pt]
    \parbox[b][\imgcellH][c]{\glabW}{\centering\rotatebox{90}{\small BG7 surface}} &
    \includegraphics[width=\gcellW]{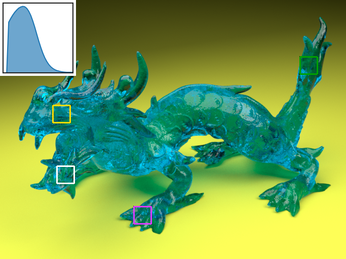} &
    \includegraphics[width=\gcellW]{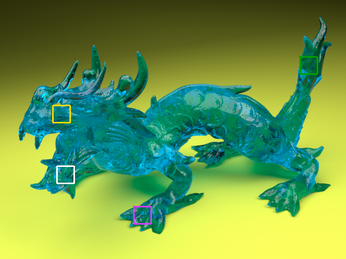} &
    \includegraphics[width=\gcellW]{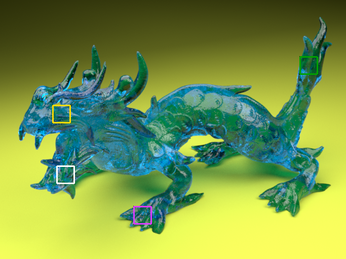} &
    \includegraphics[height=\imgcellH]{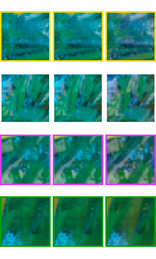} &
    \includegraphics[width=\gcellW]{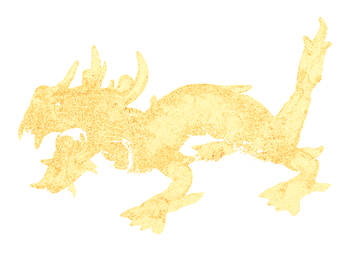} &
    \includegraphics[width=\gcellW]{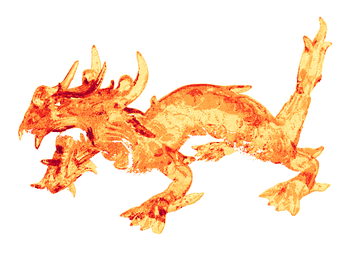} &
    \includegraphics[height=\imgcellH]{figures/material_classes/colorbar_main.png} \\[1pt]
    \parbox[b][\imgcellH][c]{\glabW}{\centering\rotatebox{90}{\small yellow-green (dipole)}} &
    \includegraphics[width=\gcellW]{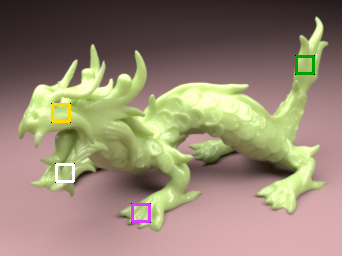} &
    \includegraphics[width=\gcellW]{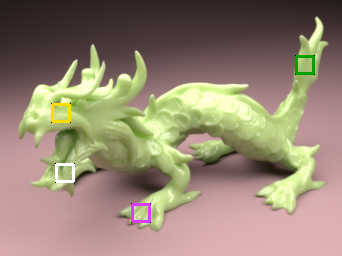} &
    \includegraphics[width=\gcellW]{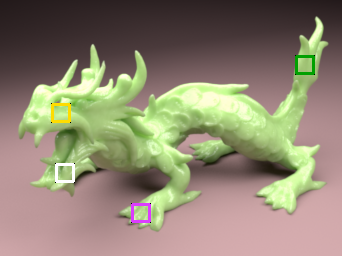} &
    \includegraphics[height=\imgcellH]{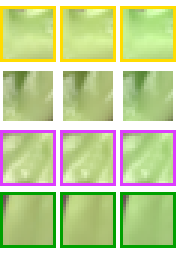} &
    \includegraphics[width=\gcellW]{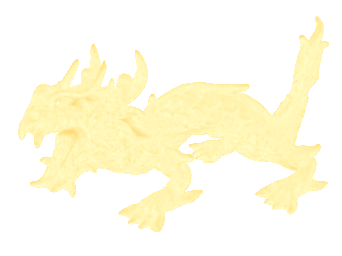} &
    \includegraphics[width=\gcellW]{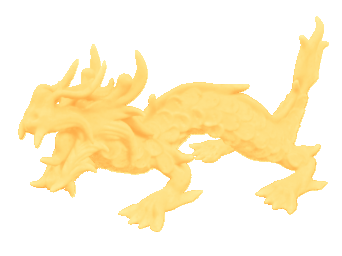} &
    \includegraphics[height=\imgcellH]{figures/material_classes/colorbar_main.png} \\[1pt]
    \parbox[b][\imgcellH][c]{\glabW}{\centering\rotatebox{90}{\small skin (homog.)}} &
    \includegraphics[width=\gcellW]{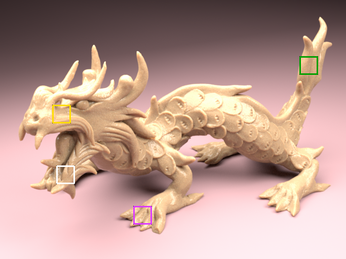} &
    \includegraphics[width=\gcellW]{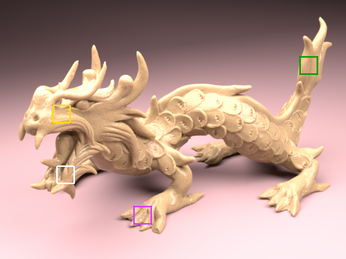} &
    \includegraphics[width=\gcellW]{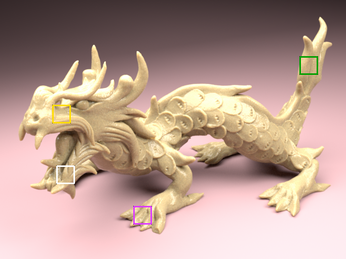} &
    \includegraphics[height=\imgcellH]{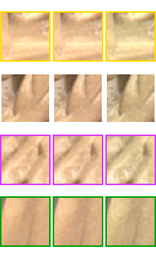} &
    \includegraphics[width=\gcellW]{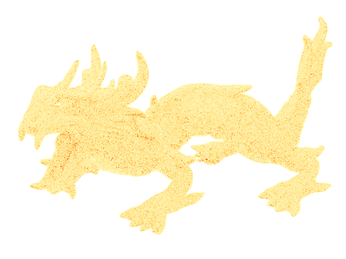} &
    \includegraphics[width=\gcellW]{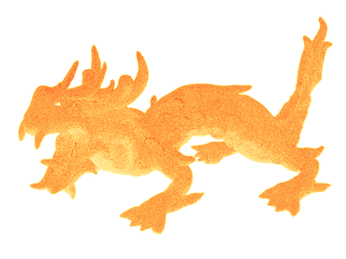} &
    \includegraphics[height=\imgcellH]{figures/material_classes/colorbar_main.png} \\[1pt]
    \parbox[b][\imgcellH][c]{\glabW}{\centering\rotatebox{90}{\small wine (homog.)}} &
    \includegraphics[width=\gcellW]{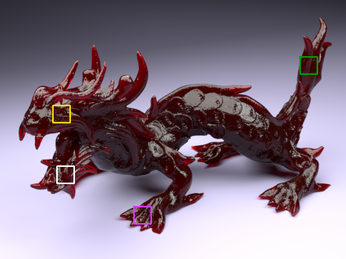} &
    \includegraphics[width=\gcellW]{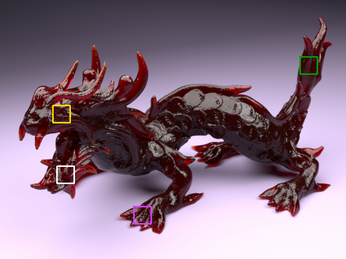} &
    \includegraphics[width=\gcellW]{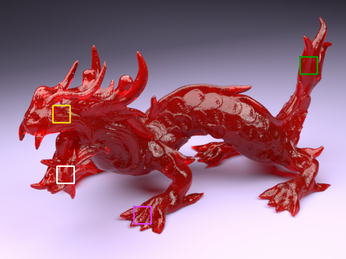} &
    \includegraphics[height=\imgcellH]{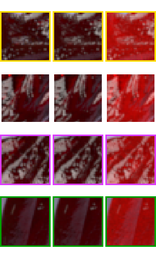} &
    \includegraphics[width=\gcellW]{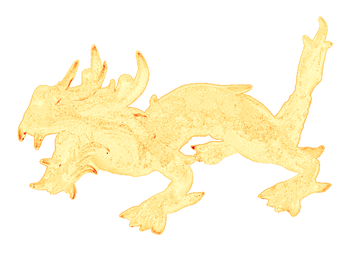} &
    \includegraphics[width=\gcellW]{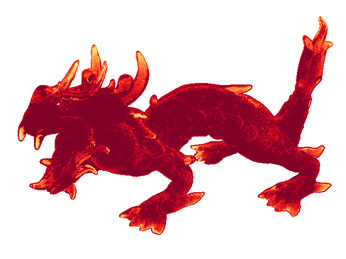} &
    \includegraphics[height=\imgcellH]{figures/material_classes/colorbar_main.png} \\
  \end{tabular}
  \caption{\textbf{Material-class evaluation.}  Seven exemplars
  spanning four regimes: glass under Beer--Lambert (B--L) and
  surface-transmittance, dipole BSSRDF, and homogeneous medium.
  Columns: spectral GT (Mitsuba\,0.6 spectral build,
  CMF$\to$sRGB); ours at $k{=}6$ (two RGB passes); RGB baseline;
  per-pixel $\Delta E_{2000}$ for the latent and the RGB baseline.
  The patches column shows four magnified regions (marked in the
  renderings; top to bottom: head, foreleg, forefoot, tail) as
  adjacent ours\,|\,GT\,|\,RGB triplets. For the glass rows the
  ground-truth cell embeds the filter's normalised transmittance
  spectrum (inset, top-left; 400--700\,nm).
  Per-material means are given in \secref{sec:material_classes};
  the remaining five materials are shown in Supplemental~Sec.~6.4.
  1024\,spp for the glass and dipole rows, 4096\,spp for the
  homogeneous rows.}
  \label{fig:mc_main}
\end{figure*}

\paragraph{Coloured glass.}
We evaluate five Schott filters (OG550, RG610, and GG495 longpass;
BG7 and VG9 bandpass) in both regimes (Supplemental~Sec.~6 renders
the set as hollow vases; Fig.~\ref{fig:mc_main} shows OG550, BG7,
and RG610): a thin transmittance filter on the dragon surface, and
Beer--Lambert absorption through the solid body. Tab.~\ref{tab:glass_dragon}
reports dragon-masked $\Delta E_{2000}$ at 1024\,spp.

\begin{table}[H]
\centering
\caption{\textbf{Coloured glass.} Dragon-masked mean $\Delta E_{2000}$ at 1024\,spp for five Schott filters in the surface-transmittance and Beer--Lambert regimes; RGB baseline, latent pipeline, and their ratio.}
\label{tab:glass_dragon}
\small
\begin{tabular}{llrrr}
\toprule
regime & glass & RGB & latent & gain \\
\midrule
surface       & OG550 & $10.52$ & $1.28$ & $\mathbf{8.22\times}$ \\
              & GG495 & $3.28$  & $1.08$ & $3.02\times$ \\
              & BG7   & $6.10$  & $2.06$ & $2.96\times$ \\
              & RG610 & $3.92$  & $1.37$ & $2.85\times$ \\
              & VG9   & $1.74$  & $1.19$ & $1.46\times$ \\
\midrule
Beer--Lambert & OG550 & $10.36$ & $1.72$ & $\mathbf{6.02\times}$ \\
              & GG495 & $3.57$  & $1.51$ & $2.36\times$ \\
              & BG7   & $4.14$  & $1.86$ & $2.23\times$ \\
              & VG9   & $3.45$  & $1.62$ & $2.13\times$ \\
              & RG610 & $2.93$  & $1.80$ & $1.63\times$ \\
\bottomrule
\end{tabular}
\end{table}

OG550 shows the largest gain in both regimes: RGB cannot represent
its sharp mid-visible cutoff. The smallest gains come from VG9 and
RG610, whose spectra RGB approximates least badly.

\paragraph{Subsurface scattering.}
We compare the two subsurface regimes, dipole BSSRDF and homogeneous
medium, side by side
(\secref{sec:materials:sss}; yellow-green dipole, skin, and wine
homogeneous in Fig.~\ref{fig:mc_main}; pale mint, rose, and potato
dipole in Supplemental~Sec.~6.4). Homogeneous dragons
encode spectral $\sigma_a(\lambda), \sigma_s(\lambda)$ (a spectral
skin profile and the red-wine coefficients of the teaser scene)
through the
unmodified renderer's per-channel volume integrator; dipole dragons
encode custom chromophore-band RGB $\sigma$ targets (their names
describe the resulting appearance), smooth-upsampled to spectra and
codec-encoded. Tab.~\ref{tab:sss_dragon} reports dragon-masked $\Delta E_{2000}$ at 1024\,spp (dipole)
and 4096\,spp (homogeneous).

\begin{table}[t]
\centering
\caption{\textbf{Subsurface scattering.} Dragon-masked mean $\Delta E_{2000}$ at 1024\,spp (dipole) and 4096\,spp (homogeneous); RGB baseline, latent pipeline, and their ratio.}
\label{tab:sss_dragon}
\small
\begin{tabular}{llrrr}
\toprule
mode & material & RGB & latent & gain \\
\midrule
dipole      & yellow-green & $3.49$  & $1.09$ & $\mathbf{3.20\times}$ \\
            & pale mint    & $1.90$  & $1.52$ & $1.25\times$ \\
            & rose         & $3.97$  & $2.66$ & $1.49\times$ \\
            & potato       & $2.76$  & $1.22$ & $2.27\times$ \\
\midrule
homogeneous & skin         & $5.61$  & $2.19$ & $2.56\times$ \\
            & wine         & $14.39$ & $2.19$ & $\mathbf{6.57\times}$ \\
\bottomrule
\end{tabular}
\end{table}

The codec advantage is largest on saturated chromophore-band materials
(wine, yellow-green) and smallest on near-broadband translucents
(potato, pale mint), where the RGB baseline already captures the
appearance reasonably.

\begin{figure}[t]
  \centering
    \setlength{\tabcolsep}{1pt}
    \renewcommand{\arraystretch}{1}
    \begin{tabular}{@{}c@{\,}c@{}}
      \scriptsize Spectral GT &
      \scriptsize patches / colour bar \\[1pt]
      \includegraphics[width=\handcellW]{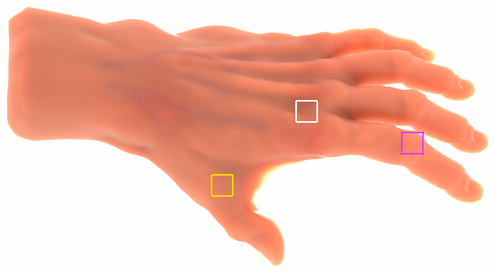} &
      \includegraphics[width=\handcellW]{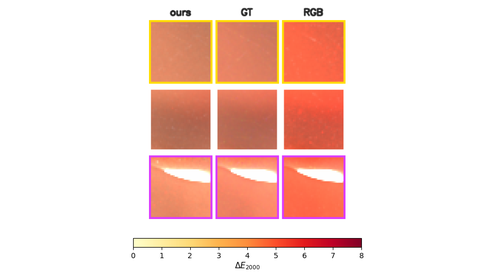} \\[3pt]
      \scriptsize Ours (latent, $k{=}6$) &
      \scriptsize $\Delta E_{2000}$ Latent \\[1pt]
      \includegraphics[width=\handcellW]{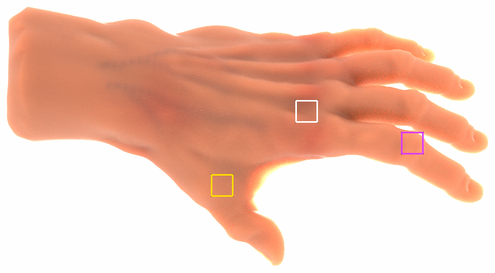} &
      \includegraphics[width=\handcellW]{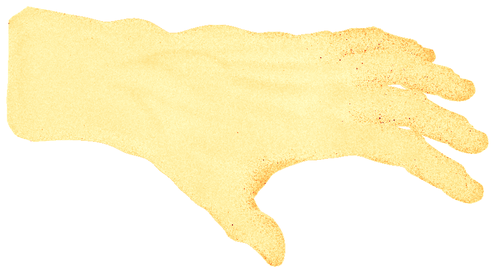} \\[3pt]
      \scriptsize RGB baseline &
      \scriptsize $\Delta E_{2000}$ RGB \\[1pt]
      \includegraphics[width=\handcellW]{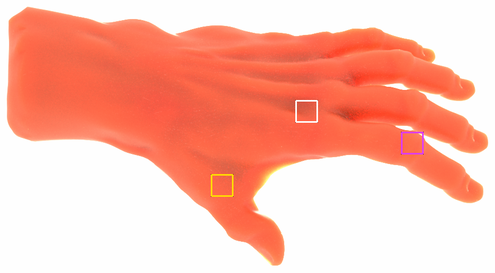} &
      \includegraphics[width=\handcellW]{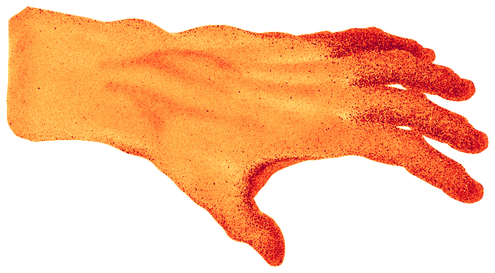} \\
    \end{tabular}
    \caption{\textbf{Heterogeneous subsurface scattering.}  Per-voxel
    spectral hand volume rendered through Mitsuba's
    heterogeneous-medium plugin; the albedo lifts five anatomical
    features to spectra via Gitlina et al.'s skin reflectance
    LUT~\shortcite{Gitlina2020}; construction details in Supplemental~Sec.~6.3.
    Each rendering sits next to its own $\Delta E_{2000}$ map; the
    top-right cell shows three magnified regions (palm, knuckle ridge,
    fingertip; marked in the renderings) as adjacent
    ours\,|\,GT\,|\,RGB triplets above the shared colour bar.
    Codec gain is $\mathbf{4.29\times}$
    (4096\,spp).}
    \label{fig:mc_hand}
\end{figure}

\paragraph{Heterogeneous subsurface scattering.}
We render a per-voxel spectral hand volume through the renderer's
heterogeneous-medium machinery (Fig.~\ref{fig:mc_hand}). The albedo
is built from five anatomical features (baseline skin, freckles,
knuckles, fingertips, veins) blended over the geometry; the warm
features are lifted to spectra via Gitlina et al.'s skin reflectance
LUT~\shortcite{Gitlina2020}, and the vein target via a chromaticity-faithful
basis fit~\cite{Guarnera2022} since it sits outside the skin gamut
(construction details in Supplemental~Sec.~6.3). At 4096\,spp the codec
attains mean $\Delta E_{2000}\!=\!0.93$ over the hand silhouette
against $4.00$ for the RGB baseline through the same volume
(p95 $1.89$ vs.\ $6.62$), a $4.29\times$ gain. The advantage spans
the whole visible surface: the RGB baseline's error map is
substantially higher across the hand, and highest on the fingers,
knuckles, and veins (Fig.~\ref{fig:mc_hand}). The heterogeneous advantage is
geometry-bounded: it grows with the
optical path length through coloured material, so it is most
pronounced on solid bodies and muted on thin shells.

\paragraph{Conductors.}
We render five measured conductors (silver, gold, copper,
aluminium, chromium) as a solid dragon under D65 (renders for the
first three in Fig.~\ref{fig:mc_metals}). Reflectance on a
conductor is a smooth but spectrally non-trivial Fresnel function of
two complex IOR spectra, which the standard RGB pipeline must
collapse to a single three-channel reflectance; the codec instead
preserves the encoded $\eta,\kappa$ pair and computes the Fresnel
response per pass (\secref{sec:materials:conductor}; renderer
bindings and a mirror-based variant in Supplemental~Sec.~6.1).
Tab.~\ref{tab:conductors_dragon}
reports dragon-masked $\Delta E_{2000}$ at 1024\,spp.

\begin{table}[H]
\centering
\caption{\textbf{Conductors.} Dragon-masked mean $\Delta E_{2000}$ at 1024\,spp for five measured metals under D65; RGB baseline, latent pipeline, and their ratio.}
\label{tab:conductors_dragon}
\small
\begin{tabular}{lrrr}
\toprule
conductor & RGB & latent & gain \\
\midrule
Ag (silver) & $8.43$ & $2.08$ & $4.06\times$ \\
Au (gold)   & $8.98$ & $0.65$ & $\mathbf{13.9\times}$ \\
Cu (copper) & $8.83$ & $1.01$ & $8.79\times$ \\
Al (aluminium) & $8.45$ & $2.11$ & $4.00\times$ \\
Cr (chromium) & $9.27$ & $2.10$ & $4.42\times$ \\
\bottomrule
\end{tabular}
\end{table}

Most of the RGB baseline's error is a magnitude error: collapsing the
complex IOR to a three-channel Fresnel overestimates conductor
reflectance by roughly a factor of two before brightness matching,
with a residual hue shift for the coloured metals (gold, copper) and
almost none for silver, whose spectral Fresnel is nearly neutral.
The $\Delta E_{2000}$ maps saturate across the entire dragon silhouette for
the RGB baseline while the latent column stays uniformly low
(Fig.~\ref{fig:mc_metals}).

\newlength{\mgcellW}\setlength{\mgcellW}{0.145\textwidth}
\newlength{\mimgcellH}\setlength{\mimgcellH}{0.1087\textwidth}
\newlength{\mglabW}\setlength{\mglabW}{0.024\textwidth}
\begin{figure*}[t!]
  \centering
  \setlength{\tabcolsep}{1pt}
  \renewcommand{\arraystretch}{1}
  \begin{tabular}{@{}c@{\,}c@{\,}c@{\,}c@{\,}c@{\,}c@{\,}c@{\,}c@{}}
    & \scriptsize Spectral GT
    & \scriptsize Ours (latent, $k{=}6$)
    & \scriptsize RGB baseline
    & \scriptsize patches
    & \scriptsize $\Delta E_{2000}$ Latent
    & \scriptsize $\Delta E_{2000}$ RGB & \\[1pt]
    \parbox[b][\mimgcellH][c]{\mglabW}{\centering\rotatebox{90}{\small Ag (silver)}} &
    \includegraphics[width=\mgcellW]{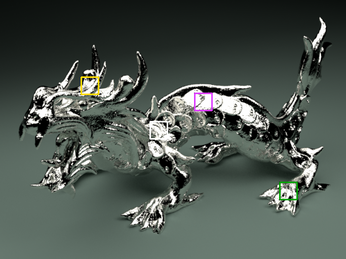} &
    \includegraphics[width=\mgcellW]{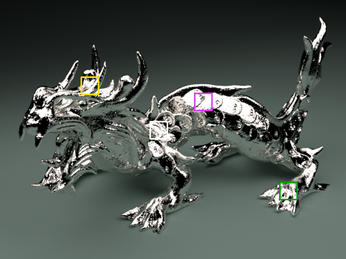} &
    \includegraphics[width=\mgcellW]{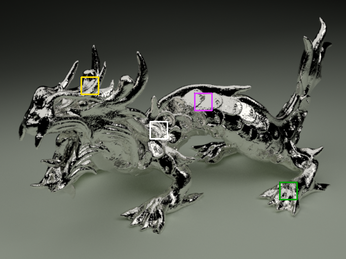} &
    \includegraphics[height=\mimgcellH]{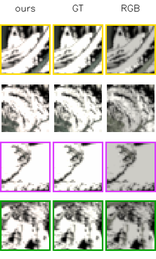} &
    \includegraphics[width=\mgcellW]{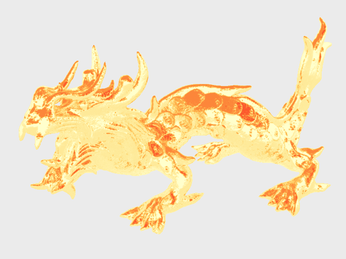} &
    \includegraphics[width=\mgcellW]{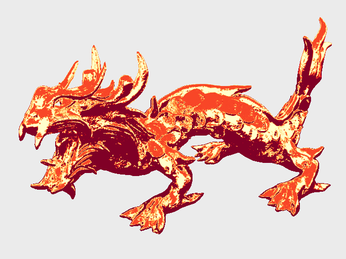} &
    \includegraphics[height=\mimgcellH]{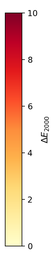} \\[1pt]
    \parbox[b][\mimgcellH][c]{\mglabW}{\centering\rotatebox{90}{\small Au (gold)}} &
    \includegraphics[width=\mgcellW]{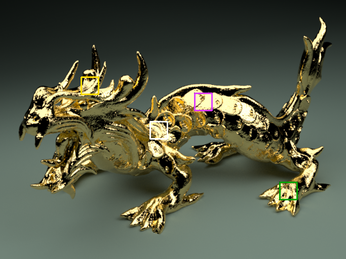} &
    \includegraphics[width=\mgcellW]{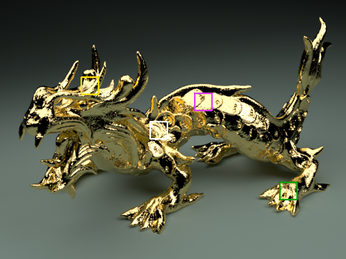} &
    \includegraphics[width=\mgcellW]{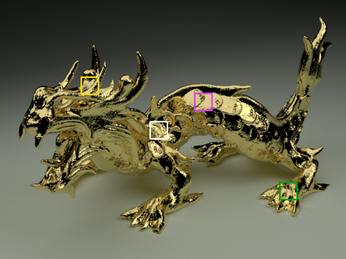} &
    \includegraphics[height=\mimgcellH]{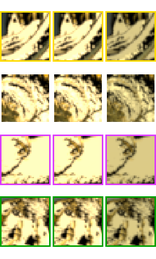} &
    \includegraphics[width=\mgcellW]{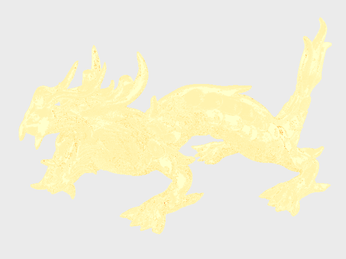} &
    \includegraphics[width=\mgcellW]{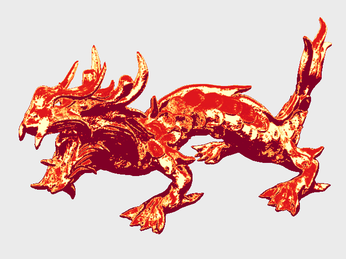} &
    \includegraphics[height=\mimgcellH]{figures/metals/colorbar_metals.png} \\[1pt]
    \parbox[b][\mimgcellH][c]{\mglabW}{\centering\rotatebox{90}{\small Cu (copper)}} &
    \includegraphics[width=\mgcellW]{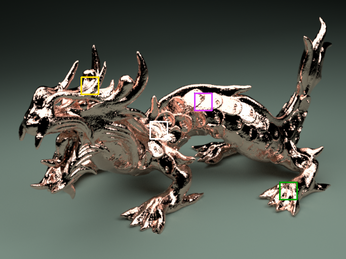} &
    \includegraphics[width=\mgcellW]{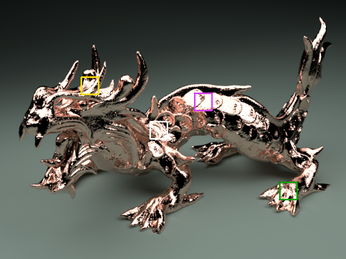} &
    \includegraphics[width=\mgcellW]{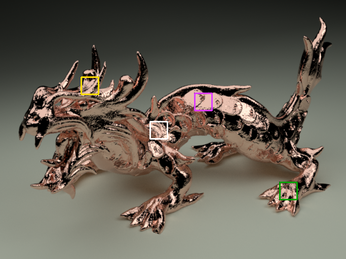} &
    \includegraphics[height=\mimgcellH]{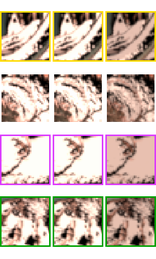} &
    \includegraphics[width=\mgcellW]{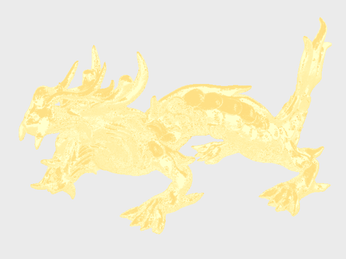} &
    \includegraphics[width=\mgcellW]{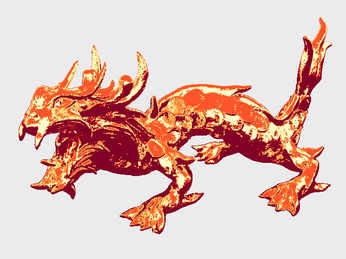} &
    \includegraphics[height=\mimgcellH]{figures/metals/colorbar_metals.png} \\
  \end{tabular}
  \caption{\textbf{Material-class evaluation: conductors.} Three
  measured conductors (silver, gold, copper) rendered as a solid
  dragon under D65, $k{=}6$ codec, dragon-masked $\Delta E_{2000}$
  versus spectral GT. Per-metal means are given in
  \secref{sec:material_classes}.
  Spectral conductor IOR is encoded via the per-component
  $\eta,\kappa$ scheme of \secref{sec:materials:conductor}. The patches column shows four magnified regions
  (marked in the renderings) as adjacent ours\,|\,GT\,|\,RGB
  triplets.}
  \label{fig:mc_metals}
\end{figure*}

\paragraph{Multi-bounce stability.}
On the Mitsuba \texttt{matpreview} test scene (Supplemental~Sec.~2.4) under
broadband and FL11 illumination, per-pixel
error stabilises immediately: from direct lighting to one bounce
the mean $\Delta E_{2000}$ rises by at most $0.02$, then remains
flat through $32$ bounces (the maximum tested) at
$0.91/1.62$ (broadband) and $1.32/1.93$ (FL11) for mean/p95.
The approximate Hadamard product does not accumulate at any depth;
the codec self-product check in Supplemental~Sec.~2.3 corroborates this.

\paragraph{Approximation behaviour.}
Each per-channel regime of \secref{sec:materials} replaces an exact
spectral operation with a latent one. The bounds that follow are
worst-case by construction: the stress chains iterate a single
encoded spectrum, whereas real paths mix different surfaces whose
encoding errors partially cancel, so rendered scene means sit at the
foot of the curves. Even so, the deviations are bounded and
slow-growing: repeated self-products of
the same encoded spectrum stay below a $\Delta E_{2000}$ of $5$
through twelve surface crossings; Beer--Lambert transmittance error
rises from the codec's product floor ($0.96$) to $1$--$6$ at the
reference depth; per-channel conductor Fresnel is nearly
angle-independent below $60^\circ$, adding about two units at
grazing incidence; and the single-scatter albedo ratio stays below
$6$ through twenty scattering orders for smooth subsurface
materials, while saturated media use the factorised encoding. The
characterisation curves, evaluated exactly as the pipeline operates
and overlaid with the rendered scene means of this section, are in
Supplemental~Sec.~6.5.

\begin{table}[H]
  \centering
  \caption{\textbf{Per-spectrum $\Delta E_{2000}$ on 218 held-out
  reflectances under four illuminants} (mean / p95), grouped by what
  each route requires of the renderer. Bold marks the best mean and
  the best p95 per column \emph{within each group}. Scaled uplift:
  the illuminant's RGB uplifted with Jakob \& Hanika and multiplied
  by D65, Mitsuba\,3's RGB environment-map convention. Guarnera et
  al.~\shortcite{Guarnera2022}: per-spectrum basis fit by
  genetic-algorithm search, with their recommended 6-LED RGBW+ basis
  or their Gaussian basis.}
  \label{tab:upsampling_dE}
  \small
  \setlength{\tabcolsep}{3pt}
  \resizebox{\columnwidth}{!}{%
  \begin{tabular}{lcccc}
    \toprule
    Method                 & D65          & LED-B1       & LED-B3       & FL11         \\
    \midrule
    \multicolumn{5}{l}{\textbf{Reflectance upsampling} (ground-truth illuminant)} \\[1pt]
    \multicolumn{5}{l}{\emph{Rendered spectrally: requires a spectral renderer}} \\
    \quad Jakob \& Hanika           & 0.65 / 1.33 & \textbf{2.00} / \textbf{8.49} & \textbf{1.39} / \textbf{4.31} & \textbf{2.01} / 6.96 \\
    \quad Guarnera et al., LED      & 0.09 / \textbf{0.03} & 5.78 / 12.62 & 3.06 / 6.09  & 2.25 / \textbf{5.32}  \\
    \quad Guarnera et al., Gaussian & \textbf{0.08} / \textbf{0.03} & 5.96 / 13.48 & 5.41 / 10.08 & 4.63 / 10.77 \\[2pt]
    \multicolumn{5}{l}{\emph{Rendered through the codec: two RGB passes}} \\
    \quad Jakob \& Hanika           & 1.59 / 2.83  & 2.57 / 8.05  & 1.89 / 6.33  & 2.91 / 5.86  \\
    \quad Guarnera et al., LED      & 2.53 / 5.45  & 4.51 / 10.70 & 3.11 / 6.60  & 3.80 / 6.92  \\
    \quad Guarnera et al., Gaussian & 1.81 / 3.66  & 4.50 / 11.32 & 3.09 / 7.73  & 5.42 / 10.46 \\
    \quad Our reflectance MLP $f^{(\mathrm{R})}$ & \textbf{0.90} / \textbf{1.74} & \textbf{1.45} / \textbf{3.88} & \textbf{1.17} / \textbf{2.61} & \textbf{1.99} / \textbf{4.56} \\
    \midrule
    \multicolumn{5}{l}{\textbf{Illumination upsampling} (ground-truth reflectance)} \\[1pt]
    \multicolumn{5}{l}{\emph{Rendered spectrally: requires a spectral renderer}} \\
    \quad Scaled uplift (J\&H$\times$D65) & \textbf{0.06} / \textbf{0.11} & \textbf{1.18} / \textbf{4.11} & \textbf{1.43} / 3.58 & 1.48 / 4.68 \\
    \quad Guarnera et al., LED      & 2.13 / 4.23  & 4.49 / 10.34 & 1.49 / \textbf{3.35}  & \textbf{1.40} / \textbf{4.40}  \\
    \quad Guarnera et al., Gaussian & 2.31 / 6.63  & 3.09 / 6.63  & 1.59 / 3.46  & 2.64 / 5.47  \\[2pt]
    \multicolumn{5}{l}{\emph{Rendered through the codec: two RGB passes}} \\
    \quad Scaled uplift (J\&H$\times$D65) & \textbf{0.97} / \textbf{1.90} & 1.49 / 4.95 & 1.59 / 5.35 & \textbf{1.70} / 4.88 \\
    \quad Guarnera et al., LED      & 3.07 / 5.73  & 4.55 / 12.12 & 1.97 / 6.17  & 1.82 / \textbf{4.34} \\
    \quad Guarnera et al., Gaussian & 2.44 / 5.82  & 3.38 / 8.52  & 2.16 / 4.66  & 2.65 / 4.88  \\
    \quad Our illumination MLP $f^{(\mathrm{L})}$ & 1.61 / 4.10 & \textbf{0.96} / \textbf{2.15} & \textbf{0.86} / \textbf{3.33} & 1.79 / 5.38 \\
    \midrule
    \multicolumn{5}{l}{\emph{Reference: no upsampling, codec on ground-truth spectra (two RGB passes)}} \\
    \quad Codec, $k{=}6$            & 0.93 / 1.85  & 1.07 / 1.91  & 1.03 / 2.18  & 1.58 / 3.17 \\
    \bottomrule
  \end{tabular}}
\end{table}

\subsection{Upsampling evaluation}
\label{sec:results_upsampling}

\paragraph{Per-spectrum upsampling accuracy.}
For each test reflectance $R$ and illuminant $L$ we compare the rendered radiance $\hat{R}\,L$ (or $R\,\hat{L}$) against the spectral ground truth $R\,L$ in CIELAB with $L$'s own integrated white point. Tab.~\ref{tab:upsampling_dE} reports $\Delta E_{2000}$ across 218 held-out reflectances and four CIE illuminants for both upsampling directions. Rows are grouped by what the route asks of the renderer, because the baselines are spectral methods: each uplifts an RGB asset to a full spectrum, which a spectral renderer then transports. That is the capability our setting assumes is unavailable, so those rows are context rather than a comparison with ours. Encoding the same uplifted spectrum and rendering it in two RGB passes puts every method in the pipeline we target, and it is those groups that compare like with like. The bottom row applies the codec directly to ground-truth spectra and serves as the no-upsampling reference.

Four observations follow from Tab.~\ref{tab:upsampling_dE}. First,
on the reflectance side the uplifting baselines are the most accurate
under D65, which is their construction point: each is fitted to
reproduce the target colour under the illuminant its fit is anchored
to, and drifts as the illuminant departs from it, so their tails grow
sharply under the structured sources while the latent path stays
bounded. Second, this anchoring is most visible in the basis fit of
Guarnera et al.~\shortcite{Guarnera2022}: designed for illumination, it reproduces its
calibration point almost exactly, and the same fit applied to
reflectance drifts further as the illuminant departs from it. Third, on the illumination side the
comparison splits by illuminant: the scaled uplift is the more
accurate under D65, the two are within $0.1$ under FL11, and our
upsampler is the most accurate under both LED sources, where it also
undercuts the ground-truth-spectra reference. Fourth, uplifting to a
spectrum and then encoding composes two approximations, at a
measurable cost against predicting the code directly ($2.24$ against
$1.38$ on average); the direct route exists to avoid that composition.
The reflectance upsampler's mean signed CIELAB offset matches the
no-upsampling reference on every illuminant, and is smaller than the
uplift-then-encode route's on all four (Supplemental~Sec.~7). Robustness
beyond these four sources is probed next.

\paragraph{Measured lamps.}
To probe illumination robustness beyond the four CIE illuminants of
Tab.~\ref{tab:upsampling_dE}, we evaluate the same per-spectrum
pipeline on 48 measured commercial and standard lamps (LSPDD
compact-fluorescent, metal-halide, and high-pressure sodium
sources \cite{LSPDD2026} plus the CIE FL and HP series, none in the
training corpus). Tab.~\ref{tab:lamps} separates two questions. First,
the representation itself holds up on unseen spectra: encoded directly,
the codec stays close to its in-distribution accuracy across the whole
population, while a pipeline with no spectral representation is several
times less accurate. Second, recovery from RGB alone is a harder
problem: the worst cases are the structured discharge lamps, and it is
there that the routes separate, the latent upsampler having the lower
error on all 48 lamps against the basis fit of Guarnera et al.\ and on
37 of 48 against the scaled uplift, the exceptions being fluorescent
standards where the two are comparable. The same population tests the
reflectance side: across the 48 lamps the direct route is more
accurate than uplift-then-encode on every one.

\begin{table}[H]
  \centering
  \caption{\textbf{Illumination robustness on unseen lamps}: mean
  $\Delta E_{2000}$ over the 218 held-out reflectances, per lamp,
  aggregated over the 48 measured lamps as mean, median, and worst lamp.
  Top row: reference, the codec applied to the ground-truth lamp
  spectra, so no recovery is involved. Middle block: methods recovering
  the lamp from its RGB alone, each encoded and rendered through the
  codec as deployed, so all four rows share one route and one renderer
  requirement; bold marks the best per column within this block. Their
  spectral behaviour, evaluated as their authors intended, is in
  Tab.~\ref{tab:upsampling_dE}.
  Bottom: the RGB baseline with no spectral representation.}
  \label{tab:lamps}
  \small
  \setlength{\tabcolsep}{6pt}
  \resizebox{\columnwidth}{!}{%
  \begin{tabular}{lrrr}
    \toprule
    & \multicolumn{3}{c}{48 measured lamps} \\
    \cmidrule(lr){2-4}
    Method & mean & median & worst \\
    \midrule
    \multicolumn{4}{l}{\emph{Reference: codec on the ground-truth lamp spectra (two RGB passes)}} \\
    \quad Codec, $k{=}6$    & 1.23 & 0.98 & 3.1 \\
    \midrule
    \multicolumn{4}{l}{\emph{Lamp recovered from its RGB, then rendered through the codec (two RGB passes)}} \\
    \quad Scaled uplift (J\&H$\times$D65) & 2.98 & 2.09 & 26.3 \\
    \quad Guarnera et al., LED    & 4.28 & 3.45 & 19.8 \\
    \quad Guarnera et al., Gaussian & 3.46 & 2.95 & 9.9 \\
    \quad Our illumination MLP $f^{(\mathrm{L})}$ & \textbf{2.18} & \textbf{1.78} & \textbf{7.5} \\
    \midrule
    \multicolumn{4}{l}{\emph{No spectral representation}} \\
    \quad RGB baseline      & 5.85 & 4.44 & 32.5 \\
    \bottomrule
  \end{tabular}}
\end{table}

\paragraph{Spectral environment maps.}
Finally, we render the upsamplers (\secref{sec:upsampling}) in complete
scenes: Cornell boxes lit by measured hyperspectral environment
maps~\cite{Morimoto:19} (Fig.~\ref{fig:envmap_upsampling}), pairing each
baseline with the corresponding route of ours, Jakob and
Hanika~\shortcite{Jakob2019spectral} against $f^{(\mathrm{R})}$ with a
ground-truth envmap, and Guarnera et al.'s 6-LED
basis~\shortcite{Guarnera2022} against $f^{(\mathrm{L})}$ with ground-truth
materials. Per-cell values are in the figure. As in
Tab.~\ref{tab:upsampling_dE}, the two baselines are rendered spectrally and carry uplift error only, while every column of ours is the
deployable two-pass route and carries the codec's error as well.
``Codec, $k{=}6$'' applies the codec to ground-truth spectra
and bounds what upsampling can achieve in these scenes. In the two maps the
rendered routes are indistinguishable.

\newlength{\envh}\setlength{\envh}{0.115\textwidth}
\newlength{\envhi}\setlength{\envhi}{0.1208\textwidth}  
\definecolor{pairA}{RGB}{217,95,2}      
\definecolor{pairB}{RGB}{27,158,119}    
\newcommand{\PairA}[1]{{\setlength{\fboxrule}{1.2pt}\setlength{\fboxsep}{0pt}\fcolorbox{pairA}{white}{#1}}}
\newcommand{\PairB}[1]{{\setlength{\fboxrule}{1.2pt}\setlength{\fboxsep}{0pt}\fcolorbox{pairB}{white}{#1}}}
\begin{figure*}[t]
  \centering
  \setlength{\tabcolsep}{1pt}
  \renewcommand{\arraystretch}{0}
  \begin{tabular}{@{}c@{\,}c@{\,}c@{\,}c@{\,}c@{\,}c@{\,}c@{}}
     \scriptsize Spectral GT &
     {\color{pairA}\scriptsize Jakob \& Hanika} &
     {\color{pairA}\scriptsize Ours, $R$ up + $L$ enc} &
     {\color{pairB}\scriptsize Guarnera, env up + GT mat} &
     {\color{pairB}\scriptsize Ours, $R$ enc + $L$ up} &
     \scriptsize Ours, both up &
     \scriptsize Codec, $k{=}6$ \\[1pt]
     \includegraphics[height=\envhi]{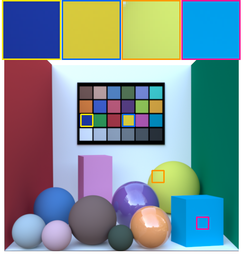} &
     \PairA{\includegraphics[height=\envhi]{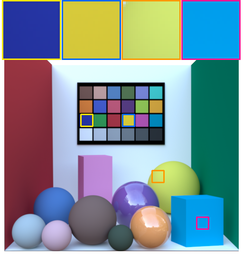}} &
     \PairA{\includegraphics[height=\envhi]{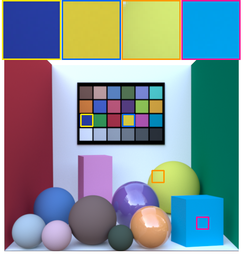}} &
     \PairB{\includegraphics[height=\envhi]{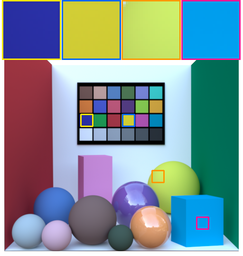}} &
     \PairB{\includegraphics[height=\envhi]{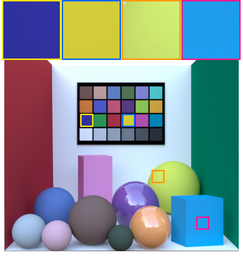}} &
     \includegraphics[height=\envhi]{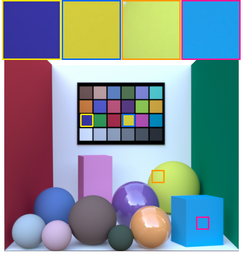} &
     \includegraphics[height=\envhi]{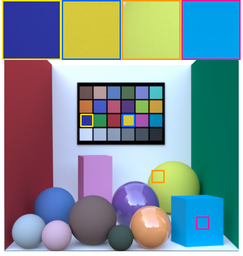} \\[1pt]
     \scriptsize Outdoor envmap &
     {\color{pairA}\scriptsize $\overline{\Delta E}\!=\!0.39$} &
     {\color{pairA}\scriptsize $\overline{\Delta E}\!=\!1.10$} &
     {\color{pairB}\scriptsize $\overline{\Delta E}\!=\!0.97$} &
     {\color{pairB}\scriptsize $\overline{\Delta E}\!=\!1.16$} &
     \scriptsize $\overline{\Delta E}\!=\!1.41$ &
     \scriptsize $\overline{\Delta E}\!=\!0.82$ \\[1pt]
     \includegraphics[height=\envh]{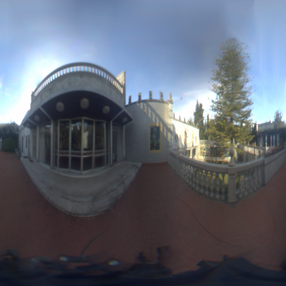} &
     \PairA{\includegraphics[height=\envh]{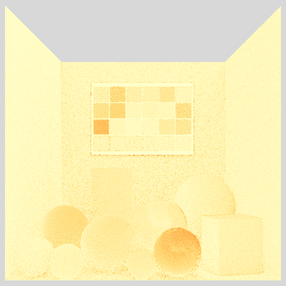}} &
     \PairA{\includegraphics[height=\envh]{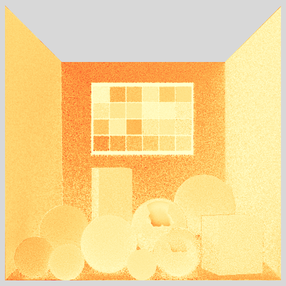}} &
     \PairB{\includegraphics[height=\envh]{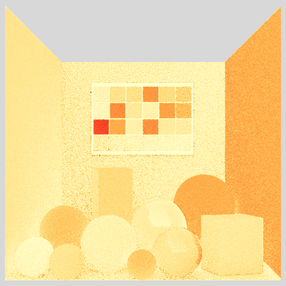}} &
     \PairB{\includegraphics[height=\envh]{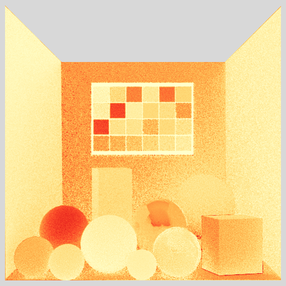}} &
     \includegraphics[height=\envh]{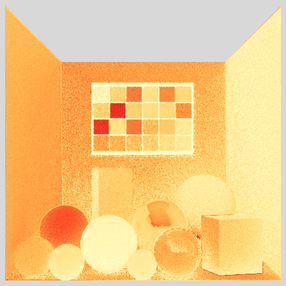} &
     \includegraphics[height=\envh]{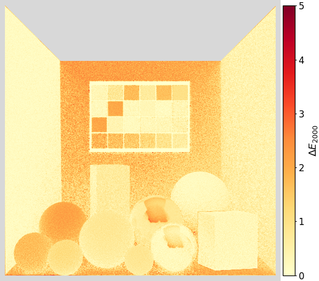} \\[5pt]
     \scriptsize Spectral GT &
     {\color{pairA}\scriptsize Jakob \& Hanika} &
     {\color{pairA}\scriptsize Ours, $R$ up + $L$ enc} &
     {\color{pairB}\scriptsize Guarnera, env up + GT mat} &
     {\color{pairB}\scriptsize Ours, $R$ enc + $L$ up} &
     \scriptsize Ours, both up &
     \scriptsize Codec, $k{=}6$ \\[1pt]
     \includegraphics[height=\envhi]{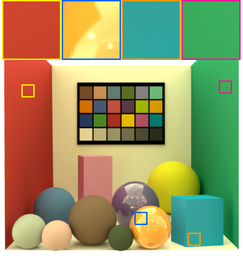} &
     \PairA{\includegraphics[height=\envhi]{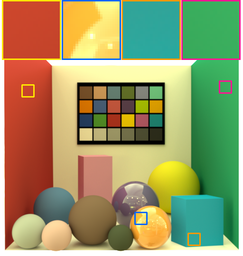}} &
     \PairA{\includegraphics[height=\envhi]{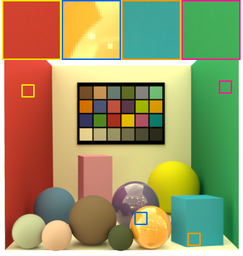}} &
     \PairB{\includegraphics[height=\envhi]{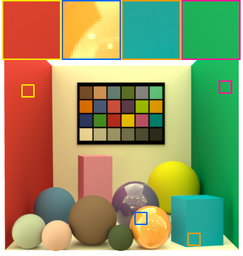}} &
     \PairB{\includegraphics[height=\envhi]{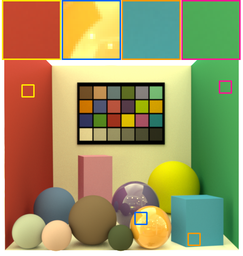}} &
     \includegraphics[height=\envhi]{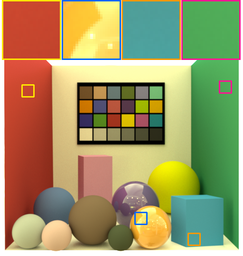} &
     \includegraphics[height=\envhi]{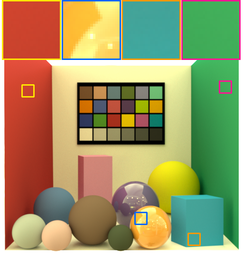} \\[1pt]
     \scriptsize Indoor envmap &
     {\color{pairA}\scriptsize $\overline{\Delta E}\!=\!0.33$} &
     {\color{pairA}\scriptsize $\overline{\Delta E}\!=\!0.41$} &
     {\color{pairB}\scriptsize $\overline{\Delta E}\!=\!0.81$} &
     {\color{pairB}\scriptsize $\overline{\Delta E}\!=\!0.81$} &
     \scriptsize $\overline{\Delta E}\!=\!0.79$ &
     \scriptsize $\overline{\Delta E}\!=\!0.32$ \\[1pt]
     \includegraphics[height=\envh]{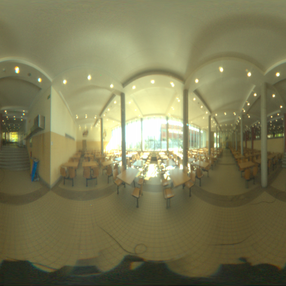} &
     \PairA{\includegraphics[height=\envh]{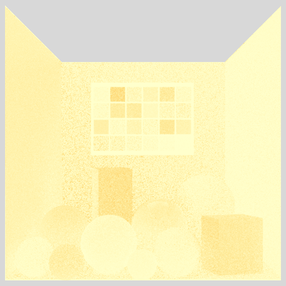}} &
     \PairA{\includegraphics[height=\envh]{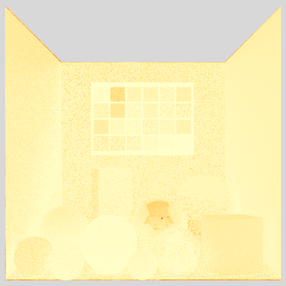}} &
     \PairB{\includegraphics[height=\envh]{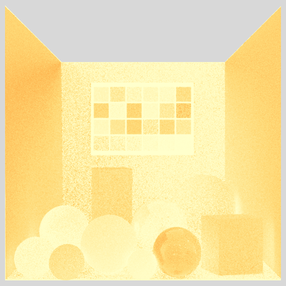}} &
     \PairB{\includegraphics[height=\envh]{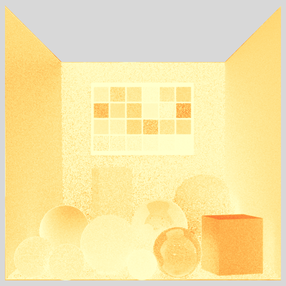}} &
     \includegraphics[height=\envh]{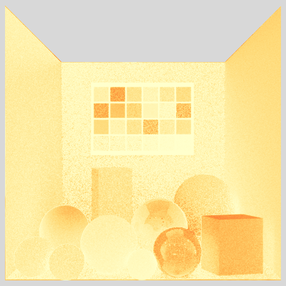} &
     \includegraphics[height=\envh]{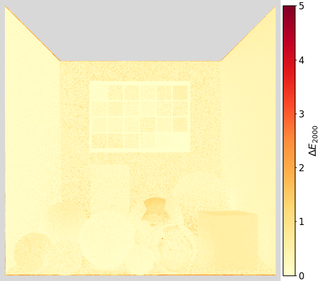} \\
  \end{tabular}
  \caption{\textbf{Latent upsampling on Cornell-box scenes lit by spectral environment maps.}  Outdoor (rows 1--2) and indoor (rows 3--4); odd rows render, even rows $\Delta E_{2000}$ with envmap thumbnail leftmost (sky pixels excluded from means).  Coloured borders group each baseline with the corresponding route of ours: \textcolor{pairA}{\textbf{reflectance upsampling with GT envmap}} (Jakob \& Hanika~\shortcite{Jakob2019spectral} vs.\ ours, $R$ up); \textcolor{pairB}{\textbf{illumination upsampling with GT materials}} (Guarnera et al.~\shortcite{Guarnera2022} vs.\ ours, $L$ up).  The two baselines are rendered spectrally and carry uplift error only; every column of ours is the deployable two-pass RGB route and carries the codec's error as well.  ``Ours, both up'' upsamples both sides from RGB; ``Ours, $k{=}6$'' applies the codec to GT spectra.  512\,spp; per-cell $\overline{\Delta E}_{2000}$ above each error map.  Hyperspectral environment maps from Morimoto et al.~\shortcite{Morimoto:19}.}
  \label{fig:envmap_upsampling}
\end{figure*}

\subsection{Cross-renderer integration}
\label{sec:cross_renderer}

To verify that the codec deploys without renderer modification, we run
it through four production frameworks: Blender Cycles (open-source path tracer;
broadband and narrowband conditions), Autodesk Maya with Arnold
(offline production), KeyShot (envmap-driven product visualisation),
and Unreal Engine 5 (real-time rasteriser with Lumen global
illumination). In each pipeline the latent code is treated as
RGB material textures and light SPDs over two passes, and the decoder
is a matrix multiply on the pass outputs.

\begin{figure*}[t]
  \centering
  \setlength{\tabcolsep}{1.5pt}
  \renewcommand{\arraystretch}{0}
  \begin{tabular}{@{}r@{\hspace{2pt}}c@{\,}c@{\,}c@{\,}c@{}}
    & \scriptsize Spectral GT (16-pass) &
    \scriptsize Codec, spectral assets &
    \scriptsize Codec + our RGB upsampler &
    \scriptsize RGB baseline \\[2pt]
    \raisebox{-0.5\height}[0pt][0pt]{\rotatebox{90}{\scriptsize\textit{narrowband illumination}}} &
    \includegraphics[width=0.2\textwidth]{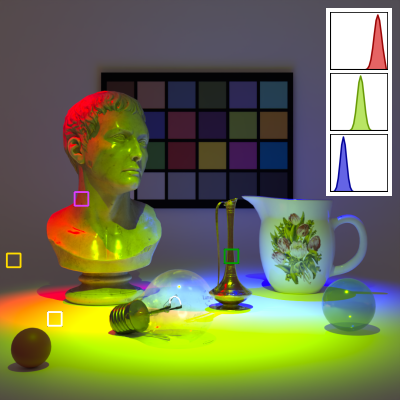} &
    \includegraphics[width=0.2\textwidth]{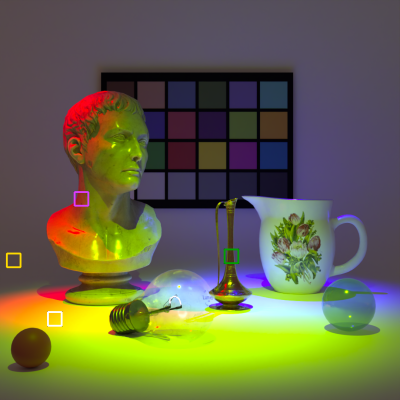} &
    \includegraphics[width=0.2\textwidth]{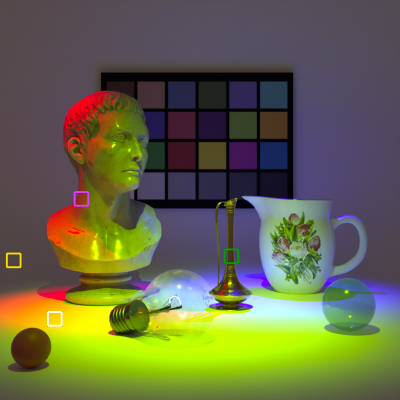} &
    \includegraphics[width=0.2\textwidth]{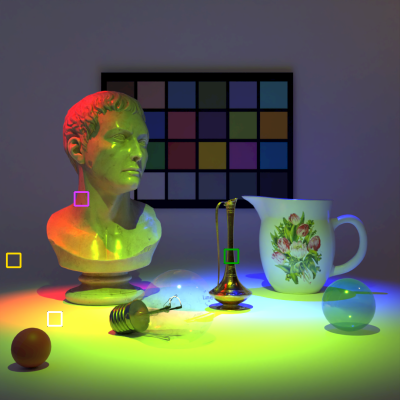} \\[1pt]
    & \includegraphics[width=0.2\textwidth]{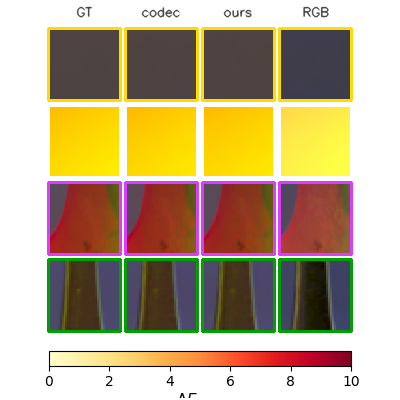} &
    \includegraphics[width=0.2\textwidth]{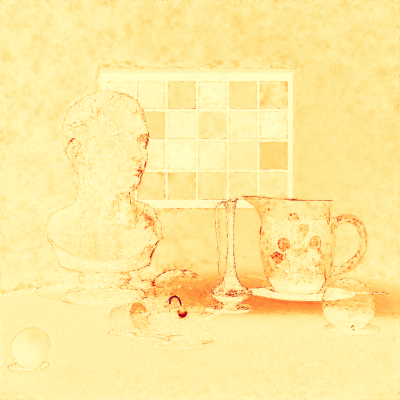} &
    \includegraphics[width=0.2\textwidth]{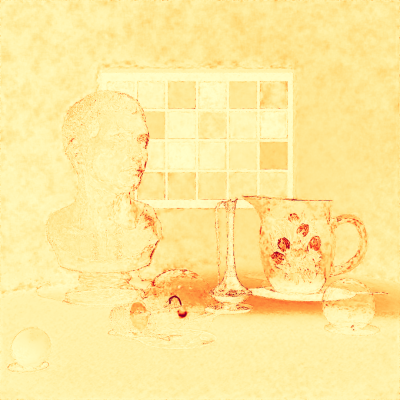} &
    \includegraphics[width=0.2\textwidth]{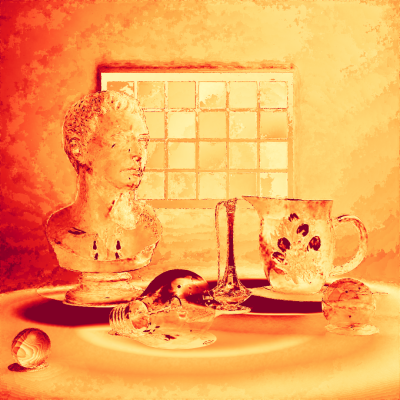} \\
    & & \scriptsize $\overline{\Delta E}\!=\!1.37$ & \scriptsize $\overline{\Delta E}\!=\!1.40$ & \scriptsize $\overline{\Delta E}\!=\!4.22$ \\[3pt]
    \raisebox{-0.5\height}[0pt][0pt]{\rotatebox{90}{\scriptsize\textit{broadband illumination}}} &
    \includegraphics[width=0.2\textwidth]{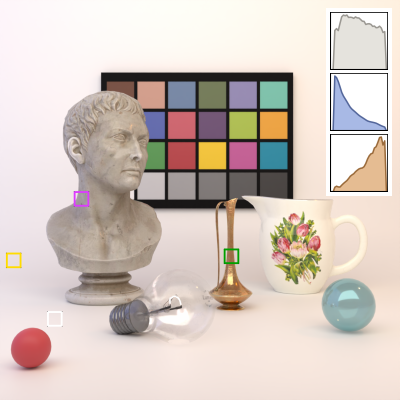} &
    \includegraphics[width=0.2\textwidth]{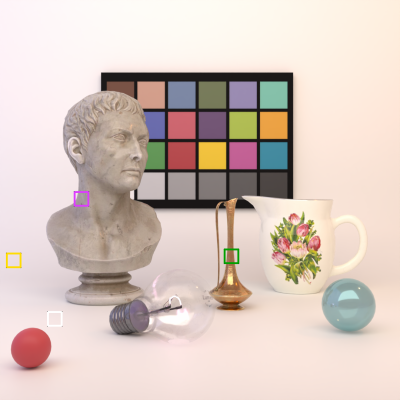} &
    \includegraphics[width=0.2\textwidth]{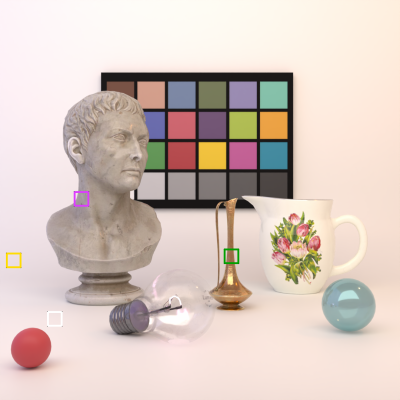} &
    \includegraphics[width=0.2\textwidth]{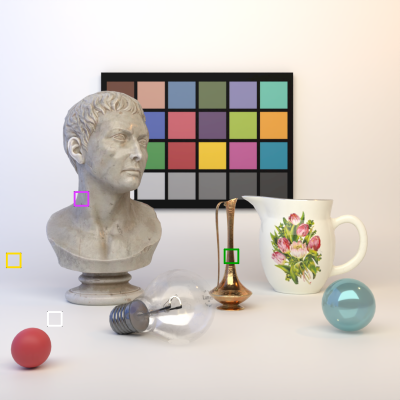} \\[1pt]
    & \includegraphics[width=0.2\textwidth]{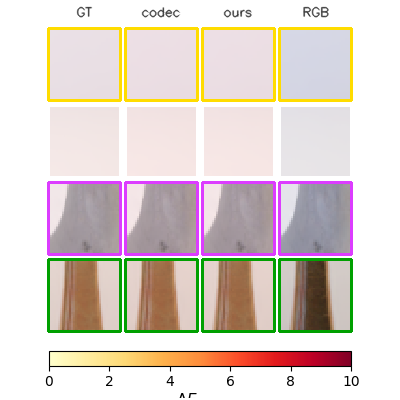} &
    \includegraphics[width=0.2\textwidth]{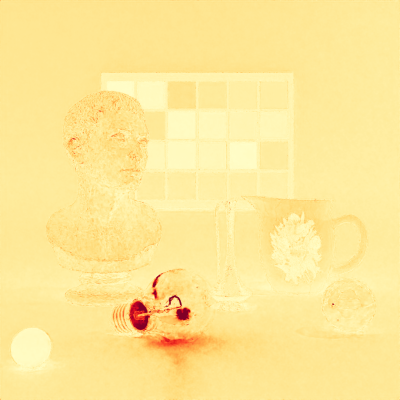} &
    \includegraphics[width=0.2\textwidth]{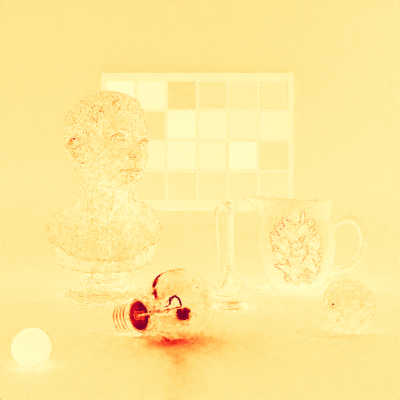} &
    \includegraphics[width=0.2\textwidth]{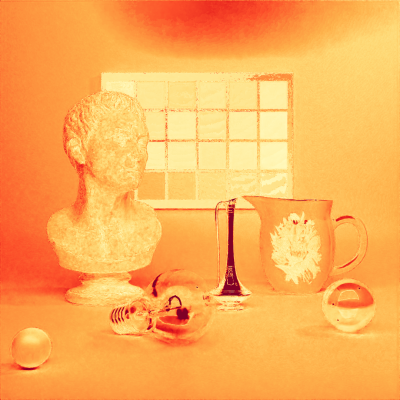} \\
    & & \scriptsize $\overline{\Delta E}\!=\!1.53$ & \scriptsize $\overline{\Delta E}\!=\!1.47$ & \scriptsize $\overline{\Delta E}\!=\!3.80$ \\[3pt]
  \end{tabular}
  \caption{\textbf{Cross-renderer evaluation: Blender Cycles} (open-source path tracer), two lighting conditions. For each lighting condition, top row: the 16-pass spectral reference; the codec applied to the same spectral assets (two RGB passes); the full RGB-asset route through our upsamplers; and the renderer's plain RGB baseline. Bottom row: four magnified patches (marked in the renderings) as adjacent GT\,|\,codec\,|\,ours\,|\,RGB swatches with the shared colour bar, then the per-pixel $\Delta E_{2000}$ of the three non-reference columns against the reference (means below; scale capped at $10$). RGB-baseline renders are brightness-matched to the reference. The inset in the reference cell shows the illumination: one panel per light SPD, or the environment map where one is used.}
  \label{fig:xr_blender}
\end{figure*}

\begin{figure*}[t]
  \centering
  \setlength{\tabcolsep}{1.5pt}
  \renewcommand{\arraystretch}{0}
  \begin{tabular}{@{}r@{\hspace{2pt}}c@{\,}c@{\,}c@{\,}c@{}}
    & \scriptsize Spectral GT (16-pass) &
    \scriptsize Codec, spectral assets &
    \scriptsize Codec + our RGB upsampler &
    \scriptsize RGB baseline \\[2pt]
    \raisebox{-0.5\height}[0pt][0pt]{\rotatebox{90}{\scriptsize\textit{light 1 (broadband)}}} &
    \includegraphics[width=0.235\textwidth]{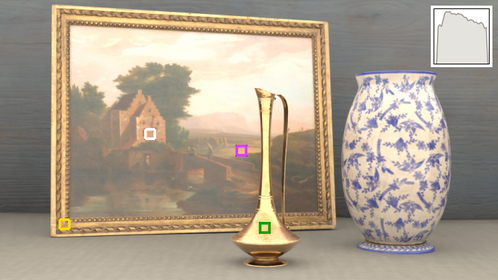} &
    \includegraphics[width=0.235\textwidth]{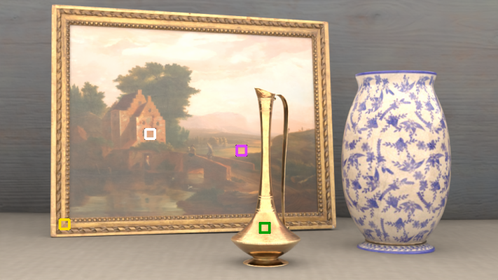} &
    \includegraphics[width=0.235\textwidth]{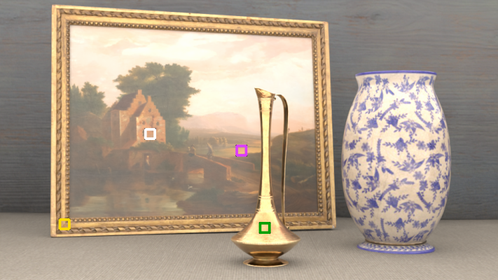} &
    \includegraphics[width=0.235\textwidth]{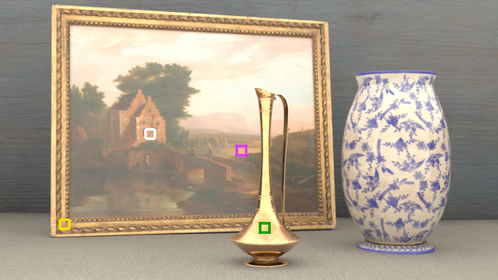} \\[1pt]
    & \includegraphics[width=0.235\textwidth]{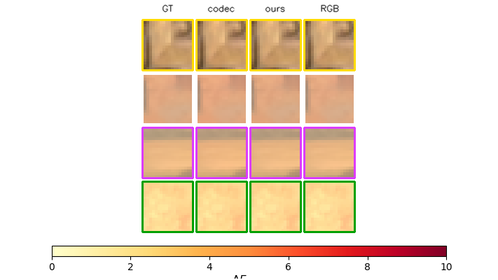} &
    \includegraphics[width=0.235\textwidth]{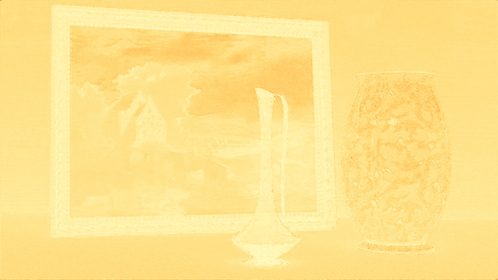} &
    \includegraphics[width=0.235\textwidth]{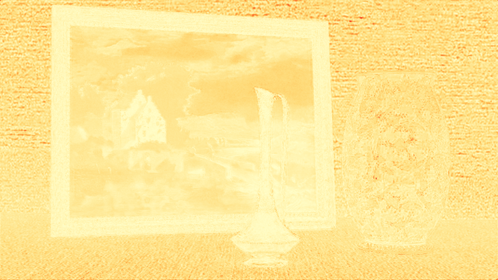} &
    \includegraphics[width=0.235\textwidth]{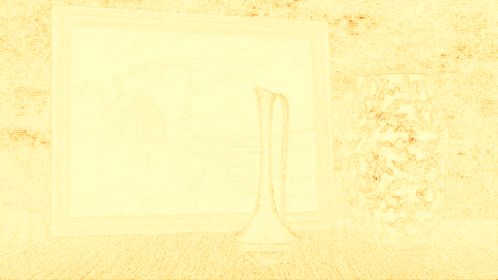} \\
    & & \scriptsize $\overline{\Delta E}\!=\!1.87$ & \scriptsize $\overline{\Delta E}\!=\!1.57$ & \scriptsize $\overline{\Delta E}\!=\!0.69$ \\[3pt]
    \raisebox{-0.5\height}[0pt][0pt]{\rotatebox{90}{\scriptsize\textit{light 2}}} &
    \includegraphics[width=0.235\textwidth]{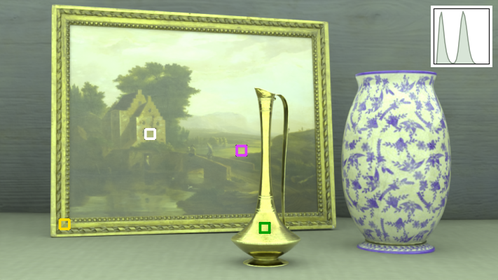} &
    \includegraphics[width=0.235\textwidth]{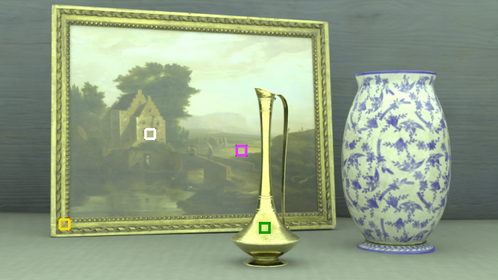} &
    \includegraphics[width=0.235\textwidth]{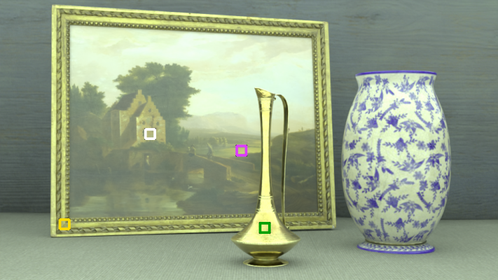} &
    \includegraphics[width=0.235\textwidth]{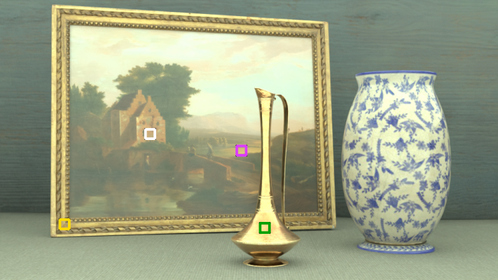} \\[1pt]
    & \includegraphics[width=0.235\textwidth]{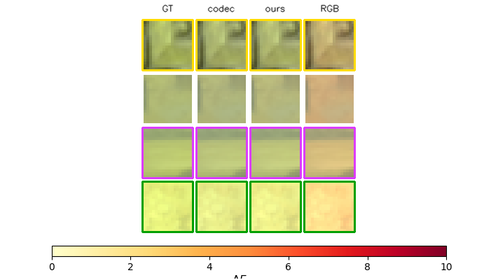} &
    \includegraphics[width=0.235\textwidth]{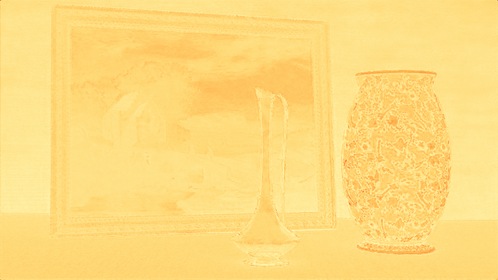} &
    \includegraphics[width=0.235\textwidth]{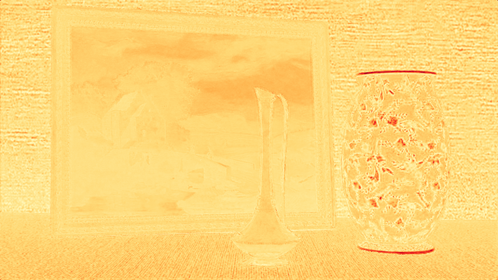} &
    \includegraphics[width=0.235\textwidth]{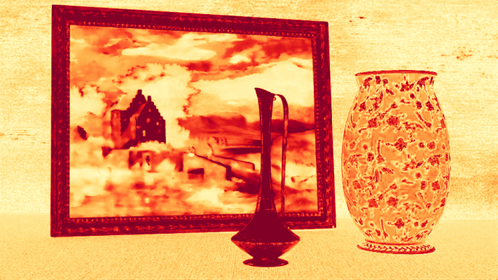} \\
    & & \scriptsize $\overline{\Delta E}\!=\!2.26$ & \scriptsize $\overline{\Delta E}\!=\!2.45$ & \scriptsize $\overline{\Delta E}\!=\!4.25$ \\[3pt]
  \end{tabular}
  \caption{\textbf{Cross-renderer evaluation: Maya Arnold} (offline production renderer), two lighting conditions: light 1 is a nearly flat broadband source, smoother than D65 and the case an RGB pipeline handles best; light 2 is a spectrally structured source. Layout as in Fig.~\ref{fig:xr_blender}. Under light 1 all three routes lie within $2\,\Delta E_{2000}$ of the reference and the RGB baseline is the most accurate ($0.69$); under light 2 the RGB baseline rises to $4.25$ while both codec routes stay near $2.3$.}
  \label{fig:xr_maya}
\end{figure*}

\begin{figure*}[t]
  \centering
  \setlength{\tabcolsep}{1.5pt}
  \renewcommand{\arraystretch}{0}
  \begin{tabular}{@{}r@{\hspace{2pt}}c@{\,}c@{\,}c@{\,}c@{}}
    & \scriptsize Spectral GT (16-pass) &
    \scriptsize Codec, spectral assets &
    \scriptsize Codec + our RGB upsampler &
    \scriptsize RGB baseline \\[2pt]
    \raisebox{-0.5\height}[0pt][0pt]{\rotatebox{90}{\scriptsize\textit{environment map}}} &
    \includegraphics[width=0.2\textwidth]{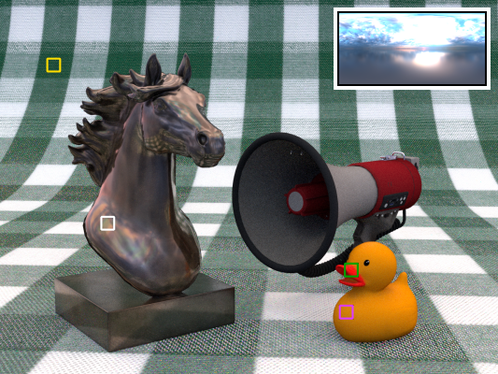} &
    \includegraphics[width=0.2\textwidth]{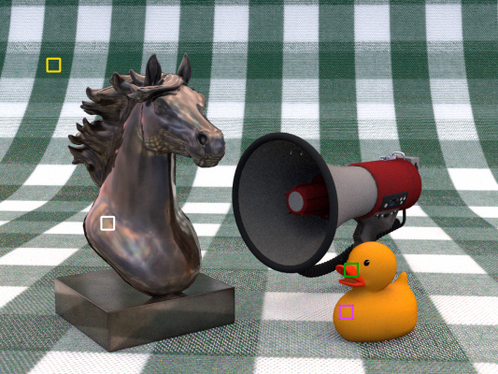} &
    \includegraphics[width=0.2\textwidth]{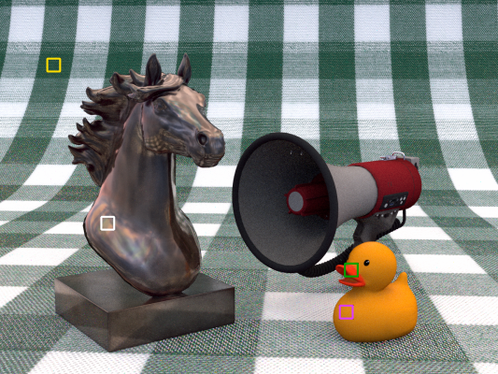} &
    \includegraphics[width=0.2\textwidth]{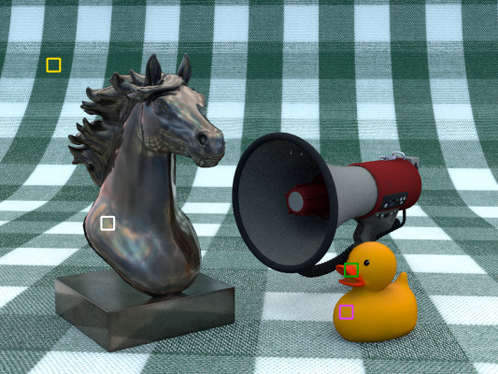} \\[1pt]
    & \includegraphics[width=0.2\textwidth]{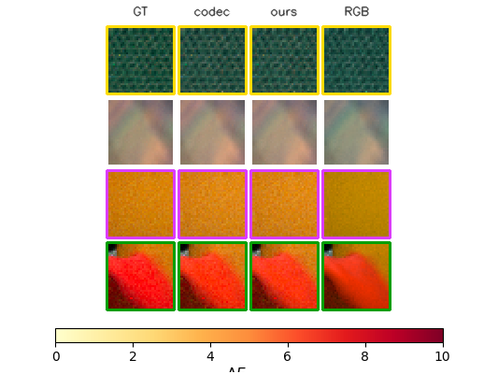} &
    \includegraphics[width=0.2\textwidth]{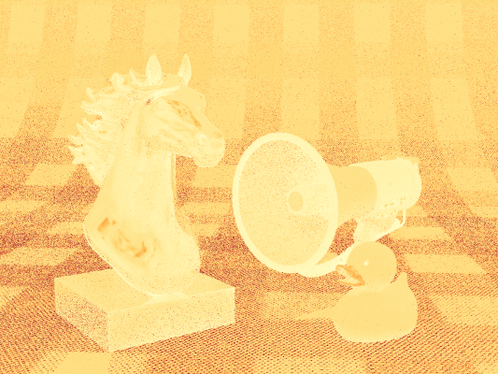} &
    \includegraphics[width=0.2\textwidth]{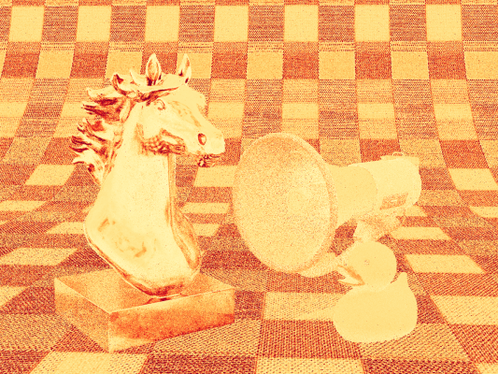} &
    \includegraphics[width=0.2\textwidth]{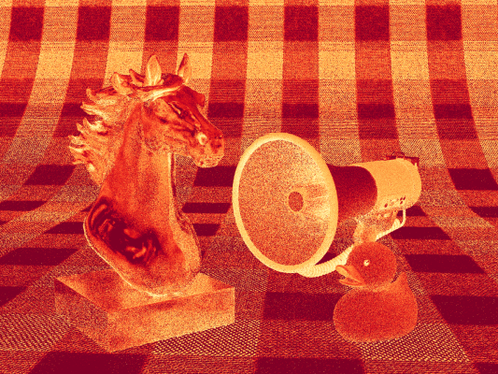} \\
    & & \scriptsize $\overline{\Delta E}\!=\!2.58$ & \scriptsize $\overline{\Delta E}\!=\!3.80$ & \scriptsize $\overline{\Delta E}\!=\!8.56$ \\[3pt]
    \raisebox{-0.5\height}[0pt][0pt]{\rotatebox{90}{\scriptsize\textit{uniform environment light}}} &
    \includegraphics[width=0.2\textwidth]{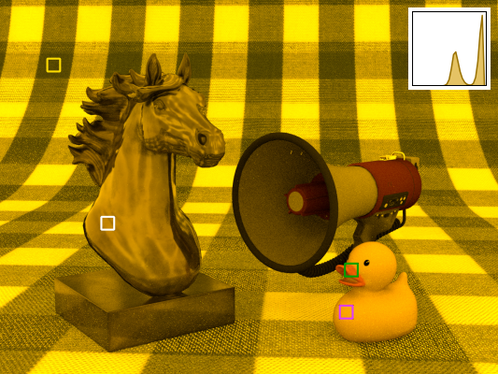} &
    \includegraphics[width=0.2\textwidth]{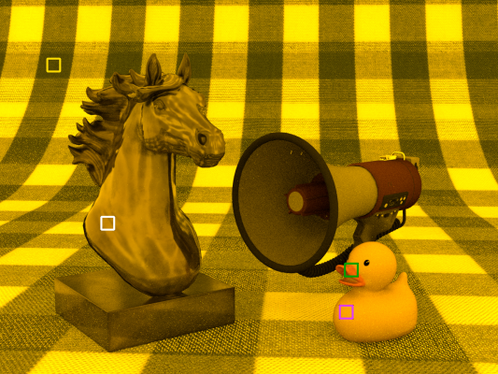} &
    \includegraphics[width=0.2\textwidth]{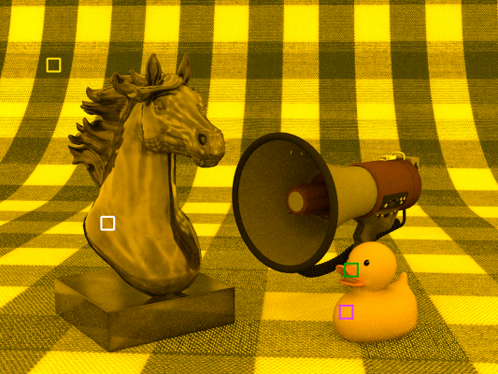} &
    \includegraphics[width=0.2\textwidth]{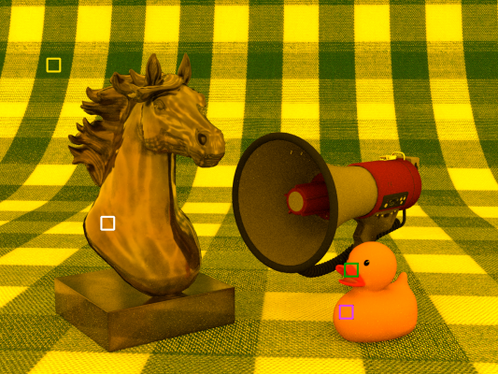} \\[1pt]
    & \includegraphics[width=0.2\textwidth]{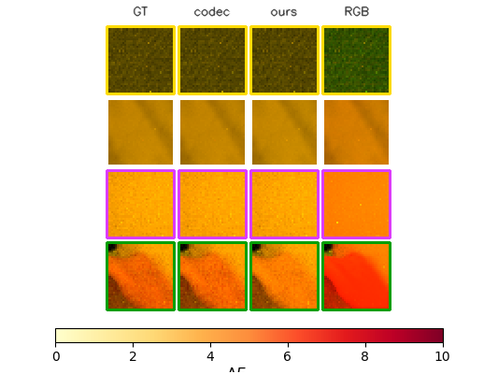} &
    \includegraphics[width=0.2\textwidth]{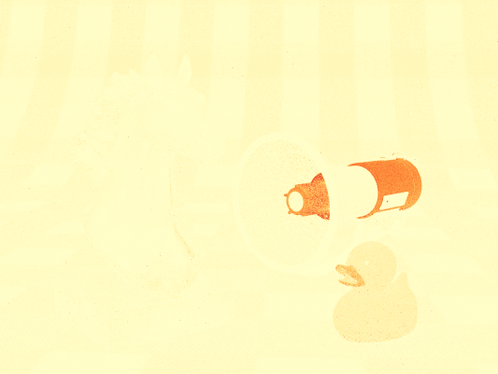} &
    \includegraphics[width=0.2\textwidth]{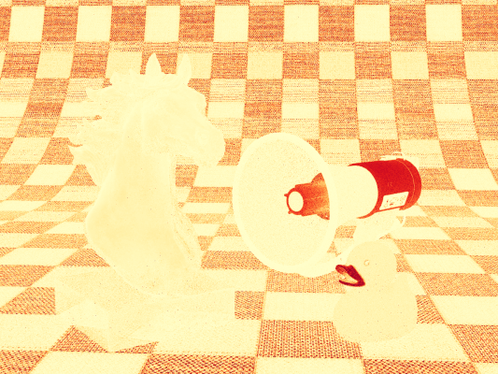} &
    \includegraphics[width=0.2\textwidth]{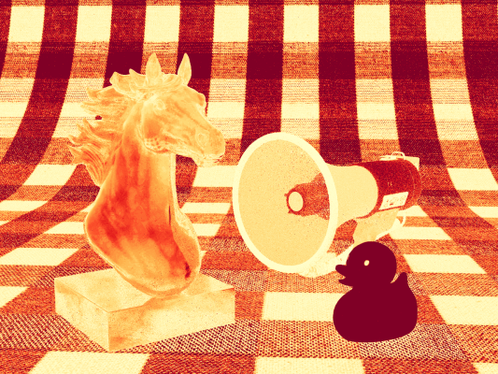} \\
    & & \scriptsize $\overline{\Delta E}\!=\!0.34$ & \scriptsize $\overline{\Delta E}\!=\!1.69$ & \scriptsize $\overline{\Delta E}\!=\!5.26$ \\[3pt]
  \end{tabular}
  \caption{\textbf{Cross-renderer evaluation: KeyShot} (product visualisation), two lighting conditions. Layout as in Fig.~\ref{fig:xr_blender}. For the environment-map condition the reference sky spectrum is the Jakob \& Hanika uplift of the RGB map, which is an 8-bit image, so the reference sky is spectrally flat wherever the map is clipped.}
  \label{fig:xr_keyshot}
\end{figure*}

\begin{figure*}[t]
  \centering
  \setlength{\tabcolsep}{1.5pt}
  \renewcommand{\arraystretch}{0}
  \begin{tabular}{@{}r@{\hspace{2pt}}c@{\,}c@{\,}c@{\,}c@{}}
    & \scriptsize Spectral GT (16-pass) &
    \scriptsize Codec, spectral assets &
    \scriptsize Codec + our RGB upsampler &
    \scriptsize RGB baseline \\[2pt]
    \raisebox{-0.5\height}[0pt][0pt]{\rotatebox{90}{\scriptsize\textit{lighting 1}}} &
    \includegraphics[width=0.235\textwidth]{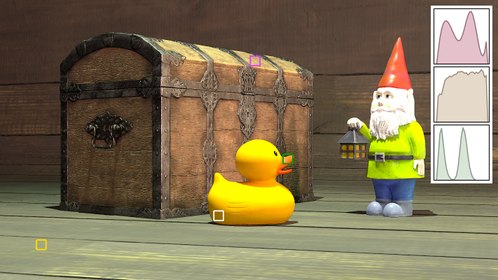} &
    \includegraphics[width=0.235\textwidth]{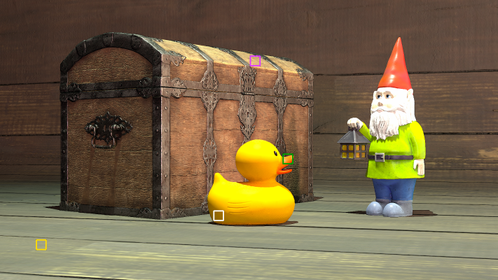} &
    \includegraphics[width=0.235\textwidth]{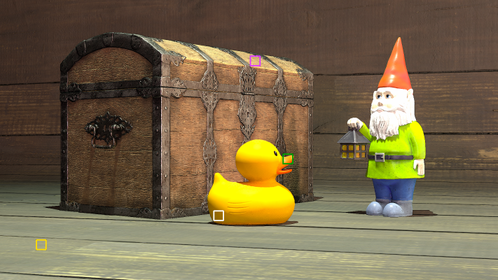} &
    \includegraphics[width=0.235\textwidth]{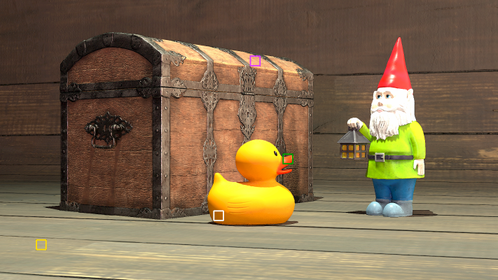} \\[1pt]
    & \includegraphics[width=0.235\textwidth]{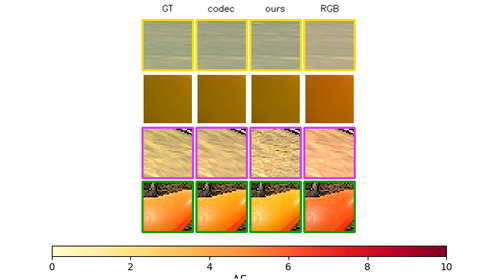} &
    \includegraphics[width=0.235\textwidth]{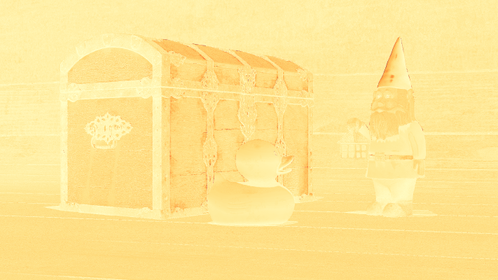} &
    \includegraphics[width=0.235\textwidth]{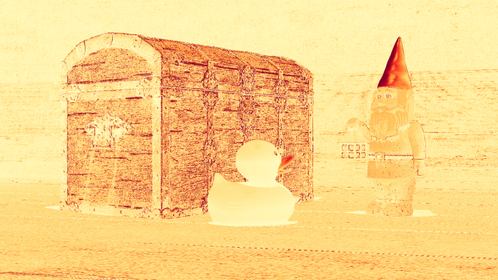} &
    \includegraphics[width=0.235\textwidth]{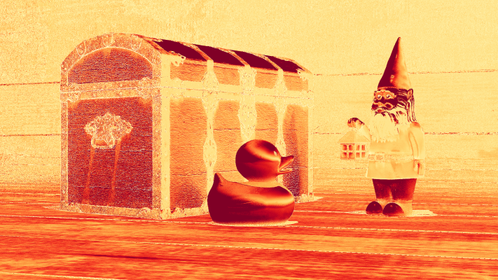} \\
    & & \scriptsize $\overline{\Delta E}\!=\!1.65$ & \scriptsize $\overline{\Delta E}\!=\!2.36$ & \scriptsize $\overline{\Delta E}\!=\!4.20$ \\[3pt]
    \raisebox{-0.5\height}[0pt][0pt]{\rotatebox{90}{\scriptsize\textit{lighting 2}}} &
    \includegraphics[width=0.235\textwidth]{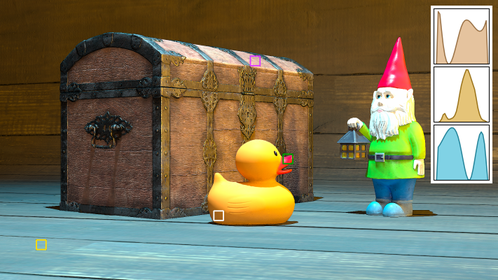} &
    \includegraphics[width=0.235\textwidth]{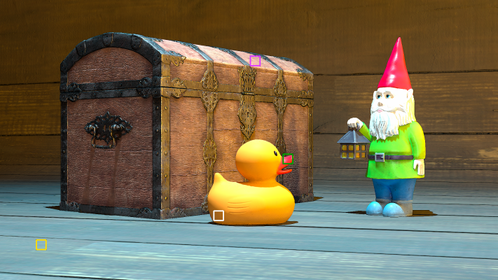} &
    \includegraphics[width=0.235\textwidth]{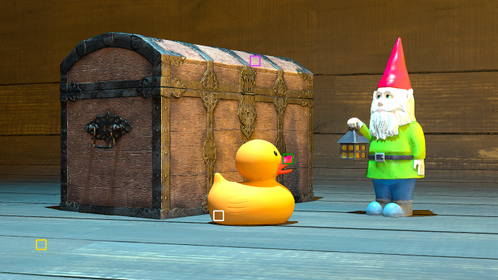} &
    \includegraphics[width=0.235\textwidth]{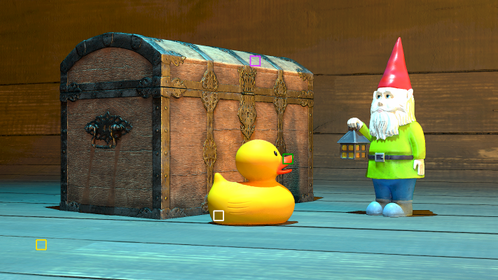} \\[1pt]
    & \includegraphics[width=0.235\textwidth]{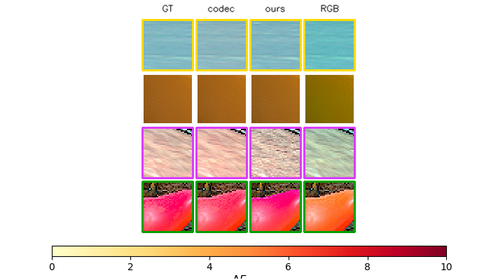} &
    \includegraphics[width=0.235\textwidth]{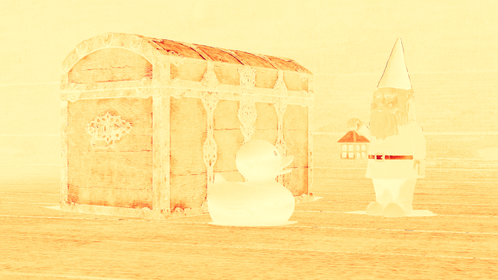} &
    \includegraphics[width=0.235\textwidth]{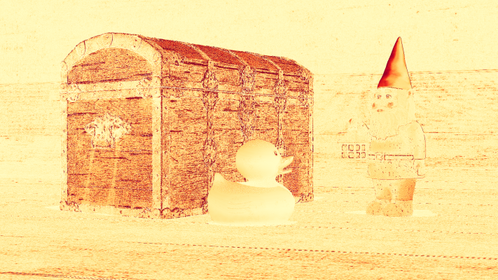} &
    \includegraphics[width=0.235\textwidth]{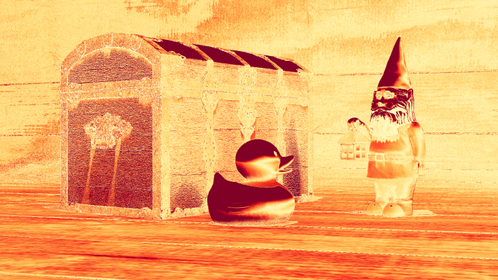} \\
    & & \scriptsize $\overline{\Delta E}\!=\!1.59$ & \scriptsize $\overline{\Delta E}\!=\!2.08$ & \scriptsize $\overline{\Delta E}\!=\!4.22$ \\[3pt]
  \end{tabular}
  \caption{\textbf{Cross-renderer evaluation: Unreal Engine 5} (real-time rasteriser with Lumen), two lighting conditions. Layout as in Fig.~\ref{fig:xr_blender}.}
  \label{fig:xr_unreal}
\end{figure*}

Spectral ground truth is not natively available in these commercial
pipelines, so we construct one per renderer: every material is
uplifted to a spectrum with Jakob and
Hanika~\shortcite{Jakob2019spectral} (the reference is not built
with our codec), light sources use their known SPDs, and each scene
is rendered in 16 RGB-packed passes that reconstruct the full spectrum, stored as linear floating-point data with
per-renderer response and exposure calibration (KeyShot and Cycles
at 128\,spp, Arnold at AA level~3, Unreal via the Movie Render
Queue; configurations in Supplemental~Sec.~9). Against this reference we
evaluate three routes per scene (Figs.~\ref{fig:xr_blender}--\ref{fig:xr_unreal}): the
codec applied to the same spectral assets in two RGB passes, which
isolates the codec's own error in each pipeline; the full
RGB-asset route through our upsamplers, which adds the recovery
cost; and the renderer's plain RGB baseline.

Across the thirteen scene and lighting cases the codec's own error stays below
$2.8\,\Delta E_{2000}$ in every renderer and lighting condition, the RGB-asset
route adds a bounded upsampling premium, largest under Unreal's
deliberately metameric lighting, and the plain RGB baseline is the
least accurate of the three in twelve of the thirteen cases. The
exception is a nearly flat broadband source in Maya, smoother across
the visible band than D65, under which all three routes fall within
$2\,\Delta E_{2000}$ of the reference. Per-case means and 95th percentiles are
in Supplemental~Sec.~9. KeyShot is evaluated both under its environment map and under
uniform-colour environment lights carrying the same known SPDs as the Maya cases; the RGB
pipelines are the renderers' own linear outputs, brightness-matched
to the reference.

A 120-frame animated sequence in Blender Cycles confirms temporal
stability: no flicker or chromatic drift across either broadband or
narrowband condition (per-renderer configurations and the
supplemental video are in Supplemental~Sec.~9).

\subsection{Performance vs.\ hero wavelength sampling}
\label{sec:results_performance}

\begin{figure*}[t]
  \centering
  \includegraphics[width=0.26\textwidth]{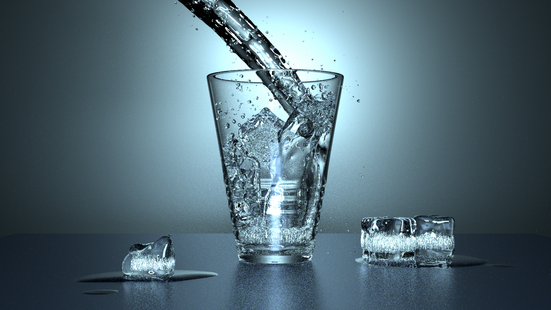}\hfill
  \includegraphics[width=0.26\textwidth]{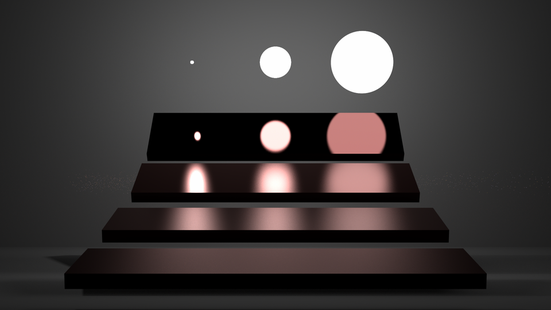}\hfill
  \includegraphics[width=0.26\textwidth]{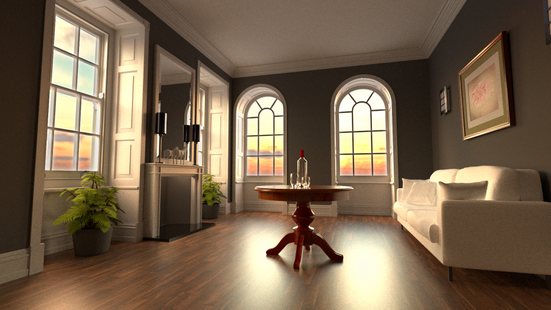}
  \caption{\textbf{Timing scenes} (spectral reference, $1920{\times}1080$,
  1024\,spp). Left to right: Glass of Water (aXel, CC0) and Veach MIS
  (B.~Bitterli, CC0), both shading-bound; the Grey \& White Room
  (Wig42, CC~BY~3.0), a textured interior whose frame time is dominated
  by traversal. The controlled-load variant adds three FL11 ceiling
  lamps outside the view.}
  \label{fig:timing_scenes}
\end{figure*}

At $k{=}6$ our method requires two RGB render passes, so its
per-frame cost is about twice an RGB render by construction and
comparable to a modern spectral renderer with hero wavelength
sampling~\cite{Wilkie2014}. We benchmark end-to-end, including codec
encode and decode, against Mitsuba\,3's spectral path tracer at
matched resolution and spp (Fig.~\ref{fig:timing_scenes},
Tab.~\ref{tab:timing_scenes}; full configurations in Supplemental~Sec.~8).

\begin{table}[H]
\centering
\caption{\textbf{Wall-clock ratio, latent (two RGB passes) over Mitsuba\,3 hero wavelength sampling}, across the spp and resolution sweep of each scene.}
\label{tab:timing_scenes}
\small
\resizebox{\columnwidth}{!}{%
\begin{tabular}{llr}
\toprule
scene & frame time bound by & latent\,/\,spectral \\
\midrule
Cornell box, 15 configurations & shading   & $0.94$--$1.10$ \\
Glass of Water        & shading           & $1.13$--$1.22$ \\
Veach MIS             & shading           & $1.13$--$1.29$ \\
Grey \& White Room    & traversal         & $1.75$--$1.81$ \\
\quad + three FL11 lamps & traversal, heavier spectra & $1.63$--$1.73$ \\
\bottomrule
\end{tabular}}
\end{table}

On the Cornell box the two-pass pipeline is at parity: within
$\pm10\%$ on 14 of 15 resolution and spp configurations, median
$+5\%$, one at $+20\%$. On heavier scenes the ratio depends on how
much of the frame time is wavelength-dependent shading rather than
geometry traversal: near parity where shading dominates, up to
$1.8\times$ on a textured interior dominated by traversal, and adding
spectral load to that same interior narrows the gap. Reconstruction
quality across the same spp and resolution sweep is reported in
Supplemental~Sec.~8. Decoding is a single matrix multiply on the framebuffer
and stays below $0.5\%$ of frame time at 1024\,spp and above.

%% file: sections/conclusion.inc
\section{Conclusion}
\label{sec:conclusion}

We introduced a learned linear non-negative spectral codec that
compresses spectra into $k$-dimensional spectral codes, exactly
preserving scaling and addition and approximately preserving the
per-channel product on which renderer shading operates. At equal
$k$, the learned basis is consistently more accurate than binning:
on the Cornell box every binning variant is about $3\times$ less
accurate, learning the encoder over fixed boxes nearly halves the
error, and jointly learning the decoder nearly halves it again
(\secref{sec:results_codec}). At $k{=}6$, two unmodified RGB
passes through any path tracer reconstruct a spectral image to a
mean $\Delta E_{2000}$ between $0.7$ and $2.7$ (median $1.8$) of
the spectral ground truth across coloured glass, conductors, dipole
and homogeneous subsurface scattering, and a per-voxel
heterogeneous subsurface-scattering hand
(\secref{sec:material_classes}); end-to-end
wall-clock time is at near parity with Mitsuba\,3's hero wavelength
sampling on shading-bound scenes and about $1.8\times$ on a
traversal-bound interior (\secref{sec:results_performance}).
Legacy RGB assets enter the same pipeline through small
XYZ-conditioned upsampling networks that add a bounded premium over
the no-upsampling reference (\secref{sec:results_upsampling}). Deployment
is renderer-agnostic: we drove four production engines, Blender
Cycles, Maya Arnold, KeyShot, and Unreal Engine 5, through the same
two-pass interface unmodified, recovering colour information the
RGB pipeline collapses. The renderer never changes; only the
meaning of its three channels does. This is efficient \emph{spectral
rendering without a spectral renderer}.

\paragraph{Scope.}
The codec preserves only what reduces to per-channel arithmetic.
Wavelength-shifting phenomena couple distinct wavelengths during
transport and cannot be captured by a per-channel $\had$ regardless
of $k$: fluorescence, phosphorescence, and the chromatic dispersion
that arises from a wavelength-dependent index of refraction. BSDF
non-linearities that
escape per-channel arithmetic (e.g.\ a coated layer whose
internal Fresnel contains $\exp(-\sigma\,d)$ along a
wavelength-dependent path) similarly fall outside the supported
regime; \secref{sec:materials} enumerates the cases that are
inside it. Layered materials, whose cross-layer coupling is
non-linear in the spectra, are a natural target for future work
under the same constraint. Residual error is largest under extremely peaked
illuminants and after many indirect interactions, but
\secref{sec:material_classes} shows it stabilises with bounce
count rather than diverging. The advantage is concentrated where an
RGB pipeline discards spectral information. On spectrally bland
media and under smooth broadband light the RGB baseline loses
little, and the codec's own reconstruction floor of one to two
$\Delta E_{2000}$ units can match or exceed it, as in the broadband Maya case of
\secref{sec:cross_renderer}. For effects outside the regime,
the codec is a pre-encoder for the linear part of the pipeline
and would need to be paired with a renderer-side handler for the
non-linear residual.

Three practical boundaries follow from the construction. The codec
operates on the 400--700\,nm band; energy outside it contributes
little to CMF-integrated colour, and extending the wavelength grid
requires only retraining on a wider corpus, not an architectural
change. The two passes must remain linear until decode: tone
mapping or other non-linear frame operations on the
per-pass buffers before decoding would break the product algebra,
so decoding must precede any such post-processing. Finally, the
encoder weights wavelengths non-uniformly, so
near-monochromatic sources at adjacent wavelengths can encode
at different magnitudes; accuracy on measured and held-out lamps is
unaffected, and the effect is confined to near-monochromatic
light (Supplemental~Sec.~5).